\documentclass[a4paper, 12pt]{article}

\pdfoutput=1

\usepackage[english]{babel}
\usepackage{cite}
\usepackage{a4wide}
\usepackage{amsmath}
\usepackage[charter,cal=cmcal,greekuppercase=italicized]{mathdesign}
\usepackage{dsfont}
\usepackage[T1]{fontenc}
\usepackage{upgreek}
\usepackage{slashed}
\usepackage{setspace}
\usepackage{graphicx}
\usepackage{xcolor}
\usepackage{soul}
\usepackage[final,          % override "draft" which means "do nothing"
            colorlinks,     % rather than outlining them in boxes
            linkcolor=purple, % override truly awful colour choices
            citecolor=teal, %   (ditto)
            urlcolor=blue  %   (ditto)
            ]{hyperref}

\makeatletter
\DeclareRobustCommand*{\bfseries}{%
  \not@math@alphabet\bfseries\mathbf
  \fontseries\bfdefault\selectfont
  \boldmath
}
\makeatother

\newcommand{\Ll}{\ensuremath{\mathrm{L}}}
\newcommand{\Rr}{\ensuremath{\mathrm{R}}}
\newcommand{\tq}{\ensuremath{\mathrm{t}}}
\newcommand{\bq}{\ensuremath{\mathrm{b}}}

\newcommand{\dq}{\ensuremath{\mathrm{d}}}
\newcommand{\uq}{\ensuremath{\mathrm{u}}}

\newcommand{\SU}{\ensuremath{\mathrm{SU}}}
\newcommand{\U}{\ensuremath{\mathrm{U}}}

\newcommand{\CP}{\ensuremath{\mathrm{CP}}}

\newcommand{\GeV}{\ensuremath{\mathrm{GeV}}}
\newcommand{\TeV}{\ensuremath{\mathrm{TeV}}}

\newcommand{\hc}{\ensuremath{\text{h.\,c.\ }}}

\renewcommand{\Re}{\ensuremath{\operatorname{Re}}}

\newcommand{\vev}{\emph{vev}}
\newcommand{\ie}{i.\,e.~}

\newcommand{\eg}{e.\,g.~}

\usepackage[font=small]{caption}
\usepackage{soul}

\begin{document}

\onehalfspacing

\begin{titlepage}

\vspace*{-15mm}
\begin{flushright}
DESY 16-114
\end{flushright}
\vspace*{0.7cm}

\begin{center} {
\bfseries\LARGE
A new view on vacuum stability in the MSSM}
\\[8mm]
Wolfgang~Gregor~Hollik
\footnote{E-mail: \texttt{w.hollik@desy.de}}
\\[1mm]
\end{center}
\vspace*{0.50cm}
\centerline{\itshape
Deutsches Elektronen-Synchrotron (DESY)}
\centerline{\itshape
Notkestra\ss{}e 85, D-22607 Hamburg, Germany}
\vspace*{1.20cm}

\begin{abstract}
\noindent
A consistent theoretical description of physics at high energies
requires an assessment of vacuum stability in either the Standard Model
or any extension of it. Especially supersymmetric extensions allow for
several vacua and the choice of the desired electroweak one gives strong
constraints on the parameter space. As the general parameter space in
the Minimal Supersymmetric Standard Model is huge, any severe constraint
on it unrelated to direct phenomenological observations enhances the
predictability of the model. We perform an updated analysis of possible
charge and color breaking minima without relying on fixed directions in
field space that minimize certain terms in the potential (known as
``\(D\)-flat'' directions). Concerning the cosmological stability of
false vacua, we argue that there are always directions in configuration
space which lead to very short-lived vacua and therefore such exclusions
are strict. In addition to existing strong constraints on the parameter
space, we find even stronger constraints extending the field space
compared to previous analyses and combine those constraints with
predictions for the light CP-even Higgs mass in the Minimal
Supersymmetric Standard Model. Low masses for supersymmetric partners
are excluded from vacuum stability in combination with the \(125\,\GeV\)
Higgs and the allowed parameter space opens at a few \(\TeV\).
\end{abstract}
\end{titlepage}

\setcounter{footnote}{0}

\section{Introduction}

The Standard Model (SM) of particle physics is completed with the final
discovery of the Higgs boson (the SM scalar)~\cite{Aad:2012tfa,
  Chatrchyan:2012ufa} which shows the expected properties in the
experiment~\cite{Khachatryan:2016vau} and only leaves small room for
deviations from the SM predictions. However, this discovery finalized a
set of problems within the SM from which one is the hierarchy problem of
the Higgs mass~\cite{Weinberg:1975gm, Weinberg:1979bn, Gildener:1976ai,
  Susskind:1978ms} another one the discussion about the cosmological
stability of the electroweak ground state~\cite{EliasMiro:2011aa,
  Degrassi:2012ry, Zoller:2012cv, Chetyrkin:2013wya, Branchina:2013jra,
  Branchina:2014usa, Kniehl:2015nwa, DiVita:2015bha,
  DiLuzio:2015iua}. Surprisingly, the most popular extension of the SM
to solve the hierarchy problem simultaneously cures the stability
problem, which is the Minimal Supersymmetric Standard Model
(MSSM). Besides the well-known solution of the hierarchy problem by the
existence of bosonic degrees of freedom that cancel loop contributions,
similar contributions render the effective potential stable---besides
the property of the MSSM having an intrinsically stable Higgs potential
at the tree-level. This solution to all problems, however, comes along
with a bunch of new problems from which a prominent one in connection to
the stability of the vacuum state is the possible destabilization of the
Higgs potential by additional scalar degrees of freedom. Finally, the
true vacuum of the theory is related to the absolute ground state of the
scalar potential which is not exclusively dedicated to vacuum
expectation values (\vev{}s) of Higgs scalars anymore but can be due to
\vev{}s of the additional scalars that break electric and/or color
charge and/or additionally baryon and lepton number. While spontaneous
breaking of lepton number may be a desired solution to the origin of
neutrino masses~\cite{Mohapatra:1979ia, Chikashige:1980qk,
  Schechter:1981cv, Kuchimanchi:1993jg}, the spontaneous breakdown of
good gauge symmetries in the SM should be avoided in a way that
\(\SU(3)_c \times \U(1)_\text{em}\) stays intact.

It was already noticed in the early 1980s~\cite{Frere:1983ag} that
supersymmetric models tend to have charge breaking minima and in the
following rather strong constraints on the soft breaking terms have been
derived~\cite{Gunion:1987qv, Drees:1985ie, Komatsu:1988mt,
  Langacker:1994bc, Strumia:1996pr, Casas:1995pd}. Subsequently, many
attempts have been performed to improve such kind of bounds using
several optimization criteria~\cite{LeMouel:2001sf}, higher loop
effects~\cite{LeMouel:1997tk, Ferreira:2000hg, Ferreira:2001tk,
  Ferreira:2004yg}, relaxing constraints allowing for metastable
states~\cite{Claudson:1983et, Kusenko:1996jn}, constraining flavor
violation~\cite{Casas:1996de} and applying metastability constraints on
the flavor violating bounds~\cite{Park:2010wf}. A sophisticated
collection of codes checking for non-standard tree-level minima,
improving with the one-loop effective potential and calculating
tunneling rates in presence of finite temperatures by the help of
\textsc{CosmoTransitions}~\cite{Wainwright:2011kj} is given by the
\textsc{Vevacious}
collaboration~\cite{Camargo-Molina:2013qva}. Recently, the old charge
and color breaking (CCB) constraints have been analyzed and tested in
the light of the Higgs discovery at the LHC~\cite{Chowdhury:2013dka,
  Chattopadhyay:2014gfa} with an updated tunneling
analysis~\cite{Blinov:2013fta}. An investigation of the one-loop Higgs
potential in the MSSM~\cite{Bobrowski:2014dla} reveals an interrelation
of one-loop stability constraints from the Higgs sector only and
tree-level CCB constraints including colored
directions~\cite{Hollik:2015pra}. Considerations of vacuum stability are
a widely used ingredient in studies of MSSM-like
scenarios~\cite{Altmannshofer:2012ks, Camargo-Molina:2014pwa,
  Camargo-Molina:2013sta, Altmannshofer:2014qha, Bagnaschi:2015pwa}.

A general paradigm is that charge and color breaking minima in the MSSM
most probably appear in such directions in field space where the
\(D\)-terms vanish. \(D\)-terms are the quadrilinear contributions to
the full scalar potential proportional to squared gauge couplings and
therefore always positive and always seen as to win over any negative
contribution. A first more complete and rather exhaustive analysis
taking basically all directions in field space into account was given
about twenty years ago by~\cite{Casas:1995pd}, where a full list of many
special cases had been discussed.

Still, a complete analysis of the problem that somehow resides in a
satisfactory solution is not possible. We provide a possible way to
handle the existence of non-standard vacua in the MSSM scalar potential
that follows the spirit of \cite{Casas:1995pd} and goes beyond. The
minimization procedure reduces then effectively to the optimization of
the necessary condition for the existence on non-standard vacua. This
optimization, however, is neither unique nor unambiguously to be
determined. Moreover, once the vacuum tunneling probability is
addressed, a new concern for the ``optimized'' field direction may
arise: to give the strongest bound from the vacuum metastability,
configurations are rather preferred that lead to the \emph{minimal}
tunneling time. Whether or not this requirement can be exploited in
automated computer tools may be left to the programming skills of the
developers. For the pedestrian, it appears sufficient to have a clear
analytical cut although those rules are indeed not sufficient but
necessary. This analytical cut, however, should only distinguish between
a global CCB minimum and a strictly stable ``desired'' electroweak
vacuum.

Why is a reassessment of this problem needed? Besides the complete
analysis of~\cite{Casas:1995pd} not so much has been done on the
analytical level as it is quite hard and any access lacks
generality. Since this great catalog of dangerous directions and
associated bounds on the parameters has been worked out, the greatest
further achievement is the discovery of the Higgs
boson~\cite{Chatrchyan:2012ufa, Aad:2012tfa} that appears to be very
SM-like and has (for MSSM purposes) a rather high mass of \(m_{h^0} =
125\,\GeV\) as follows from the combination of ATLAS and CMS data at
\(7\) and \(8\,\TeV\)~\cite{Aad:2015zhl}. This value requires sizeable
radiative corrections, that are known to be large in the
MSSM~\cite{Ellis:1990nz, Barbieri:1990ja}. However, the available
parameter space gets very much constraint imposing the correct Higgs
mass, even if one allows for a generous theoretical error of about
\(3\,\GeV\) in the determination of this
mass~\cite{Feng:2013tvd}. Especially, to achieve this shift a large stop
mixing is needed which conversely requires large trilinear soft SUSY
breaking couplings~\cite{Carena:2013ytb, Buchmueller:2013rsa}, assisted
maybe by a large Higgsino mass parameter. These large trilinear scalar
terms, however, unambiguously lead to CCB minima and render the desired
vacuum unstable. It is therefore necessary and important to put severe
constraints on those terms in order to assure theoretical
consistency. As long as there persists to be no discovery of any
sparticles at the LHC, inferring larger lower bounds on the sparticle
masses will also lead to possibly more stable configurations as larger
SUSY masses themselves lead to larger shifts in the Higgs
mass~\cite{Feng:2013tvd} without the need for large left-right squark
mixing. Anyhow, compressed scenarios that might be hidden in the
collider searches are likely to be in trouble with the stability bounds;
especially if they tuned~\cite{Djouadi:2016oey} in such a way to
reproduce weird signatures~\cite{ATLAS-digamma, CMS-digamma}.

We proceed in this paper as follows: after introducing the four-field
scalar potential, which is basically the necessary object to deal with
in connection to the influence on the Higgs mass, we derive a generic
exclusion bound in Section 2. The anatomy of the CCB states described by
this bound is discussed in Section 3. Finally, we conclude.

\section{The four-field scalar potential}
The MSSM in fact is a multi-scalar theory and its scalar potential is a
complicated object potentially leading to undesired configurations. The
configuration space depends on the vacuum expectation values of each
field that are the field values at the minima of the potential. The
potential in general has multiple minima where only the global one is
considered to be the true ground state of the theory. If in any case the
current electroweak vacuum we are believing to be sitting in is not the
true one, this configuration will only be stable for a certain amount of
time and due to quantum tunneling the global minimum will be reached.
Moreover, we have to take care that the potential is not unbounded from
below (constraints known as UFB, \ie \emph{unbounded from below} bounds
in the literature). Taking quantum corrections (and at least the
one-loop effective potential) into account, those will always be rescued
and the quantum potential will be bounded from below \cite{Casas:1995pd,
  Bobrowski:2014dla, Hollik:2015lwa}, whereas a new deep minimum will
appear at very large field values. Contrary to large field-valued minima
that usually come along with low tunneling rates into the true ground
state, the minima discussed in this paper are close-by roughly with
\vev{}s around the SUSY scale (few \TeV).

We are especially interested in the cross-relations of current analyses
in the MSSM Higgs sector with the formation of non-standard vacua. The
missing observational evidence for SUSY partners at all paired with a
relatively heavy SM-like Higgs requests extreme parameter
configurations. Existing analytical and semi-analytical bounds on the
parameter space from the stability of the standard electroweak vacuum
still are in agreement with what is needed to cope with the current
situation. However, as we will see, most scenarios in the
phenomenological MSSM (pMSSM) where all parameters are defined as input
values at the SUSY scale suffer from charge and color breaking minima
already at the SUSY scale (or slightly above). Moreover, the usual
argument that tunneling rates to the deeper minimum are sufficiently
small does not hold as there can be always a path in field space found
where a closer vacuum shows up and fast tunneling proceeds to rolling
down towards the final true vacuum. We shall explain this further.

Knowing the ground state of the theory means knowing the origin of
spontaneous symmetry breaking means knowing the structure of the
\emph{scalar} potential. Each non-vanishing \vev{} of fermionic or
vector component fields would in addition break Lorentz symmetry and
destroy the structure of space-time. Only the scalar part can break
inner symmetries spontaneously and in a way which keeps external
symmetries intact (not to speak about supersymmetry, but to break it
we rely on soft breaking and stay ignorant about its deeper origin). The
ground state of the theory is given by the state which minimizes the
potential energy density; therefore the relevant object is actually the
\emph{effective} potential, which at tree-level is equivalent to the
classical scalar potential. In principle, quantum (one and higher loop)
effects are calculable~\cite{Coleman:1973jx, Jackiw:1974cv} and allow
for spontaneous breaking radiatively. While the SM effective potential
can be trivially made stable at the tree-level by choosing the Higgs
self-coupling positive, the same coupling runs negative at higher
energies and renders the electroweak vacuum metastable on cosmological
scales~\cite{Degrassi:2012ry, Buttazzo:2013uya}. In multi-scalar
theories as in the MSSM, the situation is more involved already at the
tree-level; a tree-level analysis of the scalar potential will result in
regions of allowed parameters. Loop corrections are not expected to
make unstable regions more stable around the scale of the relevant
\vev{}, although purely loop-induced minima may be missed.

The MSSM scalar potential is calculated according to some simple rules
and consists of three basic contributions to which we will refer as the
soft breaking, the \(F\)-term and the \(D\)-term contribution:
\begin{equation}
V = V_\text{soft} + V_F + V_D.
\end{equation}
The soft breaking part breaks supersymmetry softly and mimics the
couplings of the superpotential plus additional scalar mass terms, where
the \(F\)-terms basically follow from the superpotential as derivatives
with respect to the scalar components
\begin{equation}
  V_F = \left|\frac{\partial \mathcal{W}}{\partial \phi}\right|^2,
\end{equation}
where the sum over all scalar degrees of freedom is implicitly assumed
to keep a plain notation. In our discussion and analysis, we consider
only the chiral supermultiplets of third generation quarks as they
couple with comparably large Yukawa couplings (as superpotential
parameters) to the Higgs sector and also their corresponding trilinear
soft SUSY breaking couplings are assumed to be large. For cleanliness and
a first understanding of the ``new'' phenomena hidden in an old setup, we
leave leptons and their superpartners out of the game as we are
primarily interested in the appearance of color breaking minima. The
inclusion of third generation (s)leptons is, however, trivial and
follows the same procedure. We then define the (reduced)
superpotential of ``our'' version of the MSSM by
\begin{equation}
\mathcal{W} = \mu\; H_\dq \cdot H_\uq + y_\tq\; H_\uq \cdot Q_\Ll \bar T_\Rr
- y_\bq\; H_\dq \cdot Q_\Ll \bar B_\Rr,
\end{equation}
where we denote the left-handed quark doublet as \(Q_\Ll = (T_\Ll,
B_\Ll)\) and the two Higgs doublets as \(H_\dq = (h_\dq^0, -h_\dq^-)\) and
\(H_\uq = (h_\uq^+, h_\uq^0)\), respectively, and the \(\SU(2)_\Ll\)-invariant
multiplication by the dot product. The \(\SU(2)_\Ll\) singlets are put
into the left-chiral supermultiplets \(\bar T_\Rr = \{ \tilde t_\Rr^*,
t_\Rr^c \}\) and \(\bar B_\Rr = \{ \tilde b_\Rr^*, b_\Rr^c \}\),
respectively.

Additionally, we have to break SUSY softly which is done in the usual
way with scalar mass terms and trilinear couplings:
\begin{equation}
\begin{aligned}
V_\text{soft} =&\; m_{H_\dq}^2 |h_\dq|^2 + m_{H_\uq}^2 |h_\uq|^2 - \left(B_\mu h_\dq \cdot
h_\uq +\hc\right) \\&+ {\tilde Q}_\Ll^* \tilde m_Q^2 \tilde Q_\Ll + {\tilde t}_\Rr^*
\tilde m_t^2 \tilde t_\Rr + {\tilde b}_\Rr^* \tilde m_b \tilde b_\Rr +
\left( A_\tq h_\uq {\tilde t}_\Ll^* \tilde t_\Rr + A_\bq h_\dq {\tilde b}_\Ll^* \tilde
  b_\Rr + \hc \right).
\end{aligned}
\end{equation}
The \(D\)-term part, finally, gives additional quadrilinear terms for
the scalar potential associated with gauge couplings,
\begin{equation}\label{eq:D-terms}
V_D = \frac{g_1^2}{2} \left( \phi^\dag \frac{\Upsilon_\phi}{2} \phi \right)^2
+ \frac{g_2^2}{2} \left( \phi^\dag \frac{\sigma}{2} \phi \right)^2
+ \frac{g_3^2}{2} \left( \phi^\dag \frac{T}{2} \phi \right)^2,
\end{equation}
with the corresponding hypercharges \(\Upsilon_\phi\), weak charges
\(\sigma\) (Pauli matrices for \(\SU(2)_\Ll\)-doublets \(\phi\)) and
color charge matrices \(T\). Again, summation over all gauge multiplets
\(\phi\) is implicitly understood.

The charged Higgs directions play no role in the forthcoming discussion
since on one hand the potential is \(\SU(2)\)-invariant and may be
always rotated into the desired shape---on the other hand, the soft SUSY
breaking terms also break \(\SU(2)\) in the squark sector (as top and
bottom squarks are treated differently and additional left-right mixing
is introduced by the \(A\)-terms). Any charge breaking Higgs \vev{} will
then be related to a color breaking squark \vev{} anyway and we shall be
able to express everything in neutral Higgs \vev{}s, \(h_\dq^0\) and
\(h_\uq^0\) (for simplicity, we drop the superscript ``\(^0\)'' in the
following), as well as stop and sbottom \vev{}s \(\tilde t\) and
\(\tilde b\).\footnote{The fields are treated as classical field values,
  \(c\)-numbers, and correspond to \vev{}s at the minima of the
  effective potential with vanishing external sources---the vacuum
  configuration.}  Finally, we have the combined
top/bottom-squark--Higgs scalar potential
\begin{equation}\label{eq:hisqpot}
\begin{aligned}
V_{\tilde q, h} =\;&
{\tilde t}_\Ll^* \left(\tilde m_Q^2 + |y_\tq h_\uq|^2\right) \tilde t_\Ll
+
{\tilde t}_\Rr^* \left(\tilde m_t^2 + |y_\tq h_\uq|^2\right) \tilde t_\Rr \\ &
+
{\tilde b}_\Ll^* \left(\tilde m_Q^2 + |y_\bq h_\dq|^2\right) \tilde b_\Ll
+
{\tilde b}_\Rr^* \left(\tilde m_b^2 + |y_\bq h_\dq|^2\right) \tilde b_\Rr \\ &
- \left[ {\tilde t_\Ll}^* \left( \mu^* y_\tq\; h^*_\dq - A_\tq h_\uq \right)
  \tilde t_\Rr + \hc \right]
- \left[ {\tilde b_\Ll}^* \left( \mu^* y_\bq\; h^*_\uq - A_\bq h_\dq \right)
  \tilde b_\Rr + \hc \right]
\\ &
+ |y_\tq|^2 |\tilde t_\Ll|^2 |\tilde t_\Rr|^2
+ |y_\bq|^2 |\tilde b_\Ll|^2 |\tilde b_\Rr|^2 \\ &
+ \frac{g_1^2}{8} \left( |h_\uq|^2 - |h_\dq|^2 + \frac{1}{3} |\tilde b_\Ll|^2
  + \frac{2}{3} |\tilde b_\Rr|^2 + \frac{1}{3} |\tilde t_\Ll|^2 -
  \frac{4}{3} |\tilde t_\Rr|^2 \right)^2 \\ & + \frac{g_2^2}{8} \left( |h_\uq|^2
  - |h_\dq|^2 + |\tilde b_\Ll|^2 - |\tilde t_\Ll|^2 \right)^2 \\ &
+ \frac{g_3^2}{8} \left(|\tilde t_\Ll|^2 - |\tilde t_\Rr|^2 + |\tilde b_\Ll|^2
- |\tilde b_\Rr|^2\right)^2
  \\ &
+ (m_{H_\uq}^2 + |\mu|^2) |h_\uq|^2 + (m_{H_\dq}^2 + |\mu|^2) |h_\dq|^2 - 2 \Re(B_\mu\;
h_\dq h_\uq).
\end{aligned}
\end{equation}
Some remarks are necessary on the structure of the scalar potential
given above and how to treat the field values and their possible
phases. In the previous honorable and groundbreaking works introducing
charge and color breaking solutions for the first
time~\cite{Frere:1983ag, Gunion:1987qv} it is correctly stated that for
potentials considered in these cases, the trilinear couplings as well as
the corresponding field \vev{}s can always be chosen real and
positive. This obvious observation, however, might be used to
overconstrain the field space and therefore underconstrain the
constraints on the involved parameters. Indeed, the potential of
Eq.~\eqref{eq:hisqpot} has some freedom in the field redefinitions;
especially, it is rephasing invariant apart from the trilinear terms and
the Higgs bilinear \(\sim B_\mu h_\dq h_\uq\). The last term is real by
construction, all the others (besides the trilinears) are absolute
squares of field values. Still, we do not have the freedom to rephase
the fields in such a way, that the trilinear terms behave in a well
defined way. In particular, the choice of all fields real and positive
is not possible!  We can, for sure, find a convention for the scalar
quarks but not anymore for the Higgs fields. We therefore allow both
\(h_\uq\) and \(h_\dq\) to vary in the positive and negative regime and
only constrain \(|\tilde t| = \alpha |\phi|\) as well as \(|\tilde b| =
\beta |\phi|\) with a certain scalar field value \(\phi\) (where we
choose \(h_\uq = \phi\)). Moreover, we set \(h_\dq = \eta \phi\) with
\(\eta\) any real number and \(\alpha\), \(\beta\) real and positive. In
case, we are considering real parameters only (not the complex MSSM),
the potential is symmetric in \(\Ll \leftrightarrow \Rr\) exchange of
left- and right-handed field labels. Setting all squark fields \(\tilde
q_\Ll = \tilde q_\Rr\) (with \(q = t,b\)) simplifies also the
\(D\)-terms in the sense, that the \(g_3^2\) contribution vanishes and
the \(g_2^2\) and \(g_1^2\) are the same in terms of the fields. The
commitment to real parameters (and fields!) nevertheless is also a
severe constraint, that may be, however, compassed by imposing global
\CP-invariance of the theory (\ie~\CP-invariance of SUSY breaking if one
refers to the \(A\)-terms). It is therefore a good assumption to
consider real fields only and just constrain the colored scalars to be
positive (as the colored potential is invariant under \(\tilde q \to -
\tilde q\)).\footnote{Complex fields in the effective potential mean
  spontaneous CP violation.}

Applying the considerations from above, we now have
\begin{equation}
\begin{aligned}
V_\phi =\;& \alpha^2 (\tilde m_Q^2 + \tilde m_t^2 + 2 y_\tq^2 \phi^2)
\phi^2 + \beta^2 (\tilde m_Q^2 + \tilde m_b^2 + 2 y_\bq^2 \eta^2 \phi^2) \phi^2
\\ &
+ \left(m_{H_\uq}^2 + \eta^2 m_{H_\dq}^2 + (1+\eta^2) |\mu|^2 -2 B_\mu \eta \right) \phi^2
\\ & - 2 \alpha^2(\mu y_\tq \eta - A_\tq) \phi^3 - 2 \beta^2 (\mu y_\bq - \eta
A_\bq) \phi^3 \\ &
(\alpha^4 y_\tq^2 + \beta^4 y_\bq^2) \phi^4
+ \frac{g_1^2 + g_2^2}{8} \left(1 - \eta^2 + \beta^2 - \alpha^2
\right)^2 \phi^4,
\end{aligned}
\end{equation}
where we applied in Eq.~\eqref{eq:hisqpot}
\begin{equation}
\begin{aligned}
h_\uq = \phi, \quad & |\tilde t_\Ll| = |\tilde t_\Rr| = |\tilde t| =
\alpha |\phi|, \\
h_\dq = \eta \phi, \quad & |\tilde b_\Ll| = |\tilde b_\Rr| = |\tilde b| =
\beta |\phi|.
\end{aligned}
\end{equation}
Rewriting finally the potential, we get
\begin{equation}\label{eq:onefieldpot}
\begin{aligned}
V_\phi =& \left(m_{H_\uq}^2 + \eta^2 m_{H_\dq}^2 + (1+\eta^2) \mu^2 -2 B_\mu \eta
+ (\alpha^2 + \beta^2) \tilde m_Q^2 + \alpha^2 \tilde m_t^2 + \beta^2 \tilde m_b^2
 \right) \phi^2
\\ &
 - 2 \left( \alpha^2(\mu y_\tq \eta - A_\tq) + \beta^2 (\mu y_\bq - \eta
A_\bq) \right) \phi^3 + ( \alpha^2 y_\tq^2 + \beta^4 y_\bq^2 ) \phi^4 \\ &
+ \left( \frac{g_1^2 + g_2^2}{8} (1 - \eta^2 + \beta^2 - \alpha^2)^2
+ 2 \alpha^2 y_\tq^2 + 2 \beta^2 y_\bq^2 \right) \phi^4
\\ \equiv\;&
M^2 \phi^2 - \mathcal{A} \phi^3 + \lambda \phi^4,
\end{aligned}
\end{equation}
with
\begin{subequations}
\begin{align}
M^2 &= \; m_{H_\uq}^2 + \eta^2 m_{H_\dq}^2 - 2 B_\mu \eta + (1+\eta^2) \mu^2
+ (\alpha^2 + \beta^2) \tilde m_Q^2 + \alpha^2 \tilde m_t^2 + \beta^2 \tilde m_b^2
   \,, \label{eq:M2} \\
\mathcal{A} &= \; 2  \alpha^2 \eta \mu y_\tq - 2 \alpha^2 A_\tq + 2 \beta^2 \mu
y_\bq - 2 \eta \beta^2 A_\bq \,, \label{eq:A} \\
\lambda &= \;  \frac{g_1^2 + g_2^2}{8} (1 - \eta^2 + \beta^2 - \alpha^2)^2
+ (2 + \alpha^2) \alpha^2 y_\tq^2 + (2 \eta^2 + \beta^2) \beta^2 y_\bq^2
\,. \label{eq:lambda}
\end{align}
\end{subequations}
Each of the effective parameters in the potential
Eq.~\eqref{eq:onefieldpot} depends implicitly on the scaling parameters,
so \(M^2 = M^2(\eta, \alpha, \beta)\), \(\mathcal{A} = \mathcal{A}(\eta,
\alpha, \beta)\) and \(\lambda = \lambda(\eta, \alpha, \beta)\). The
minimization of the one field potential is done trivially and also the
requirement for the global minimum at \(\langle\phi\rangle = 0\) is
known to be
\[M^2 > \frac{\mathcal{A}^2}{4\lambda}.\]
Knowing about the dependence on the actual field direction, this bound can
be improved as
\begin{equation}\label{eq:improvedbound}
4 \min_{\{\eta,\alpha,\beta\}} \lambda(\eta, \alpha, \beta) M^2(\eta,
\alpha, \beta) > \max_{\{\eta,\alpha,\beta\}} \left(\mathcal{A}(\eta,
  \alpha, \beta)\right)^2,
\end{equation}
or rather
\[
\min_{\{\eta,\alpha,\beta\}} \left[ 4 \lambda(\eta, \alpha, \beta)
  M^2(\eta, \alpha, \beta) - \left(\mathcal{A}(\eta, \alpha,
    \beta)\right)^2 \right] > 0.
\]
Note, that we easily recover the famous ``traditional'' CCB bound by
Fr\`ere et al.~\cite{Frere:1983ag} setting \(\eta=\beta=0\) and \(\alpha
=1\) which corresponds to the ray \(|\tilde t_\Ll| = |\tilde t_\Rr| =
|h_\uq|\) in field space: \(M^2(0,1,0) = m_{H_\uq}^2 + \mu^2 + {\tilde
  m}_Q^2 + {\tilde m}_t^2\), \(\mathcal{A}(0,1,0) = -2 A_\tq\) and
\(\lambda(0,1,0) = 3 y_\tq^2\), such that\footnote{As important side
  remark, we have to admit that we defined our soft breaking \(A\)-terms
  differently from the common SUSY literature, where the Yukawa
  couplings are been factored out (to recover the original result, one
  has to replace \(A_\tq \to y_\tq A_\tq\)).}
\begin{equation}\label{eq:tradbound}
|A_\tq|^2 < 3 y_\tq^2 \left(m_{H_\uq}^2 + \mu^2 +
{\tilde m}_L^2 + {\tilde m}_t^2\right).
\end{equation}
Similar expressions can be easily achieved for different field
directions. The specific choices have been made to make all gauge
coupling contributions in Eq.~\eqref{eq:lambda} vanish---though the
quartics from the Yukawa couplings, which are numerically much larger,
remain. There exists no real solution for \(\eta\) with non-vanishing
\(\alpha\) and/or \(\beta\) to have \(\lambda = 0\). So there will be
(large) quartics anyway, which on the other hand means that we do not
necessarily need to restrict to the \(D\)-flat condition which
explicitly forces all \(g_i^2\)-terms in the scalar potential to be
absent.

Similarly, by employing other alignments, we also recover the recently
proposed~\cite{Hollik:2015pra} \(\mu y_\bq\) bound and the corresponding
bound from the \(h_\uq\)-\(\tilde b\) \(D\)-flat direction with either
\(\eta = 0\) and \(\beta = 1\),
\begin{equation}\label{eq:ccbbot}
\frac{ \left(\mu y_\bq \right)^2 }{y_\bq^2 + (g_1^2 + g_2)^2 / 2} <
m_{H_\uq}^2 + \mu^2 + {\tilde m}_Q^2 + {\tilde m}_b^2,
\end{equation}
or \(h_\dq = \pm \sqrt{1 + \alpha^2} |h_\uq|\), corresponding to
\(|h_\dq|^2 = |h_\uq|^2 + |\tilde b|^2\), and leading to
\begin{equation}\label{eq:ccbbot2}
  \frac{\alpha^2\mu^2}{2+3\alpha^2} < (1 + \alpha^2) m_{H_\dq}^2
  + m_{H_\uq}^2 + (2 + \alpha^2) \mu^2
  \pm 2 B_\mu^2 \sqrt{1+\alpha^2}
  + \alpha^2 \left( {\tilde m}_Q^2 + {\tilde m}_b^2 \right),
\end{equation}
with \(\alpha > 0\); a reasonable fit to the numerically derived
exclusion limits can be found for \(\alpha \approx 0.8\).

In the past, many attempts have been exercised to significantly improve
the stability bound on the trilinear \(A\)-term according to
Uneq.~\eqref{eq:tradbound}. Possible replacements range from
\begin{equation}\label{eq:tradCLM}
|A_\tq|^2 < 3 y_\tq^2 \left( m_{H_\uq}^2 + {\tilde m}_Q^2 + {\tilde m_t}^2
\right),
\end{equation}
which was given (actually for \(\tq \leftrightarrow \uq\) on the first
generation \(A\)-term) by~\cite{Casas:1995pd} and improved considering
the cosmological stability of the potential through tunneling effects
by~\cite{Kusenko:1996jn} to\footnote{Uneq.~\eqref{eq:KuLaSe} is
  sometimes referred to ``empirical'' bound in contrast to the
  ``traditional'' one of Uneq.~\eqref{eq:tradbound}.}
\begin{equation}\label{eq:KuLaSe}
A_\tq^2 / y_\tq^2 + 3 \mu^2 < 7.5 \left( {\tilde m}_Q^2 + {\tilde m_t}^2
\right),
\end{equation}
and recently updated by~\cite{Blinov:2013fta} in the light of the Higgs
discovery as
\begin{equation}\label{eq:BlinMorr}
A_\tq^2 / y_\tq^2 < 3.4 \left( {\tilde m}_Q^2 + {\tilde m}_t^2 \right) + 60
\left(m_{h_2}^2 + \mu^2 \right),
\end{equation}
which is more in agreement (numerically) with Uneq.~\eqref{eq:tradCLM} but
shall only be applied to smaller values of \(\mu\) and larger
pseudoscalar masses \(m_A\), whereas moderate \(\tan\beta\). How exactly
this ``small'', ``large'' and ``moderate'' is defined may be left to the
\emph{gusto} of the user. All in all, the bounds
\eqref{eq:tradbound}--\eqref{eq:BlinMorr} leave an undecided feeling
behind and remain open the question for a robust, roughly unique and
unambiguous constraint (which we also fail to provide).

We insist on the smaller sign (\(<\)) in Uneq.~\eqref{eq:tradbound} and
later on because the smaller or equal (\(\leq\)) includes a degenerate
vacuum with \(\langle \phi\rangle \neq 0\) which also leads to undesired
phenomenology (where we do not want to speculate about multiple
degenerate vacua as done for the SM Higgs
case~\cite{Froggatt:1995rt}). To be on the safe side, the \(<\) is
always preferred. The optimized class of conditions given in
Uneq.~\eqref{eq:improvedbound} lead in general to a more involved
interplay of different field directions that cannot be displayed in such
a nice expression like Uneq.~\eqref{eq:tradbound}.

The meaning of such bounds stayed controversial in the literature and
history. One significant improvement has been achieved by the discussion
about the stability on cosmological grounds, the question whether or not
the desired vacuum has had the possibility to decay to the true vacuum
within the life-time of the universe. However, any (semi-)analytical
constraint suffers from a distinct choice of the field configurations as
any such choice influences the tunneling rate, as well.

The main task is now to find the ``optimized'' directions, meaning
certain combinations of \(\eta\), \(\alpha\) and \(\beta\) that give
rise to the most severe bounds \`a la Uneq.~\eqref{eq:improvedbound}
leading to the deepest CCB minimum (and therefore the true vacuum of
the theory). Numerical minimization (and maximization) can be
efficiently done with many available tools. However, as we will see, the
optimized direction is not necessarily the most dangerous direction as
the former one is in certain cases related to very large field \vev{}s
accompanied with a rather high barrier between the trivial (local)
minimum at \(\langle \phi \rangle = 0\) and the true vacuum. Those
configurations are related to very large tunneling times for the
vacuum-to-vacuum transition and thus considered to be less
dangerous. There are nevertheless slightly tilted or shifted directions
in field space where the non-standard minimum lies closer and also the
barrier is more complanate and therewith easier to be reached by quantum
tunneling. Once the barrier is overcome, the true vacuum can be
approached directly.

Before we continue with the actual analysis of the (reduced) MSSM
incarnated in the full scalar potential of Eq.~\eqref{eq:hisqpot}, we
make a brief but necessary digression and discuss the issue of vacuum
tunneling.

\paragraph{Instability vs.\ metastability}

The process of finding the global minimum of a complicated potential is
hazardous, even more the interpretation of the newly found
configuration. Is the standard (local) vacuum stable against quantum
tunneling towards this preferred true vacuum---or may there even be a
path to gently roll down into the desired state? The estimate of the
tunneling rate via the so-called bounce action itself is a tricky
business, however, for a wide class of potentials a very pictorial
approximation can be used where only the position (i.e. \vev) of the
deeper minimum and the maximum in between and the height of the wall is
needed. For a thick wall separating the false from the true vacuum, a
very convenient approximation formula was provided
by~\cite{Duncan:1992ai} which is an exact solution for a triangular
shape of the potential. The difficult part lies in the calculation of
the bounce action \(B\)~\cite{Coleman:1977py}, the decay rate per unit
volume is then given by
\[\frac{\Gamma}{V} = A e^{-B},\]
where \(A\) is an undetermined amplitude factor, usually approximated by
the false \vev{} to the fourth power or the barrier height (as the
uncertainty goes into the exponent, this does not really matter). The
bounce itself depends in this approximation only on the true and the
false \vev{}, \(\phi_+\) and \(\phi_-\), respectively, the field value
of the maximum in between \(\phi_M\), and the values of the effective
potential at the false vacuum \(V_+ = V(\phi_+)\) as well as the peak
of the wall \(V_M = V(\phi_M)\). The difference \(\Delta V_+ = V_M -
V_+\) gives the height of the wall; furthermore, we define \(\Delta
\phi_+ = \phi_M - \phi_+\) and \(\Delta \phi_- = \phi_- - \phi_M\) and
have the bounce action of~\cite{Duncan:1992ai}
\begin{equation}\label{eq:bounce}
  B = \frac{ 2 \pi^2}{3} \frac{\left[ \left( \Delta \phi_+ \right)^2
      - \left(\Delta \phi_- \right)^2 \right]^2}{\Delta V_+}.
\end{equation}
Eq.~\eqref{eq:bounce} is very convenient to check the stability of a
given configuration in the reduced one-field potential without going
into the details of the non-perturbative calculation. In comparison to
the life-time of the universe, one finds metastable vacua for \(B
\gtrsim 400\), see~\cite{Kusenko:1996jn}.

\begin{table}
  \caption{Input values and derived quantities for the two parameters
    points illustrated in Figs.~\ref{fig:bounceample}
    and~\ref{fig:bounceample-severe}.}
  \label{tab:sample-points}
\begin{center}
  \begin{tabular}{l|llllll}
    & \(M_\text{SUSY}\) & \(\tan\beta\) & \(\mu\) & \(A_\tq = A_\bq\) & \(m_{h^0}\)
    & \(B_\text{global}\)
    \\ \hline
    Fig.~\ref{fig:bounceample} & \(1\,\TeV\) & \(40\)
    & \(500\,\GeV\) & \(1500\,\GeV\) & \(126\,\GeV\) & \(354\) \\
    Fig.~\ref{fig:bounceample-severe} & \(1\,\TeV\) & \(10\) & \(500\,\GeV\)
    & \(500\,\GeV\) & \(113\,\GeV\) & \(2568\)
  \end{tabular}
\end{center}
\hrule
\end{table}

It is not necessarily the global minimum that determines the tunneling
rate to a non-standard minimum. Numerical procedures may overlook the
vacuum on one hand, but on the other hand the decay time to a local
minimum may be much smaller and the transition to the deeper one does
not play a role anymore.\footnote{Quantum mechanics knows about all
  paths.} We want to illustrate at two sample points with different
phenomenology that both show deeper charge and color breaking
minima. The first point accounts for the proper Higgs mass with
\(m_{h^0} \approx 126\,\GeV\) where the other one would be discarded
because it has \(m_{h^0} \approx 113\,\GeV\). However, the nature of the
global minimum is different for both points: while the first has a
short-lived electroweak vacuum with \(B \lesssim 400\), the other has an
extremely long-lived false vacuum. All relevant parameters are given in
Tab.~\ref{tab:sample-points}. In all our analyses, we keep the
pseudoscalar heavy to comply with the recent exclusions by collider
searches for \(A,H \to \tau\bar\tau\)~\cite{Aad:2014vgg,
  Khachatryan:2014wca} and take \(m_A = 800\,\GeV\) (which is very
borderline with the respect to the 2014 analyses up to \(\tan\beta =
40\) but unconstrained for smaller \(\tan\beta = 10\)). The pseudoscalar
mass has anyway only a mild impact on the charge and color breaking
potential as it enters via the determination of the soft SUSY breaking
Higgs masses \(m_{H_\uq}^2\), \(m_{H_\dq}^2\) and \(B_\mu\) and we can
easily set \(m_A = M_\text{SUSY}\) without changing the results. These
three mass parameters can be related and constrained demanding the Higgs
potential being bounded from below at the tree-level \emph{and}
triggering electroweak symmetry breaking such that \(h_\uq = h_\dq = 0\)
is unstable~\cite{Casas:1995pd}:
\[
m_{H_\dq}^2 + m_{H_\uq}^2 + 2 |\mu|^2 \geq 2 |B_\mu|
\geq \sqrt{\left(m_{H_\dq}^2 + |\mu|^2 \right) \left(m_{H_\uq}^2 + |\mu|^2 \right)}.
\]
As we only check for CCB minima, we do not impose this constraint in
addition; a parameter point excluded by non-vanishing squark \vev{}s is
excluded anyways. For the allowed points in the following numerical
evaluation, this consideration should be applied. Most points do not
recover the correct light CP-even Higgs mass in the MSSM, not even
within an error range of about \(\pm 3\,\GeV\). If nothing else is
quoted, we employed the latest version (2.11.3) of
\textsc{FeynHiggs}~\cite{Hahn:2013ria, Frank:2006yh, Degrassi:2002fi,
  Heinemeyer:1998np, Heinemeyer:1998yj} to determine its numerical
value. We include a discussion of the influence of a \(125\,\GeV\) Higgs
in Sec.~\ref{sec:anat}.

\begin{figure}
\begin{minipage}{0.48\textwidth}
\includegraphics[width=\textwidth]{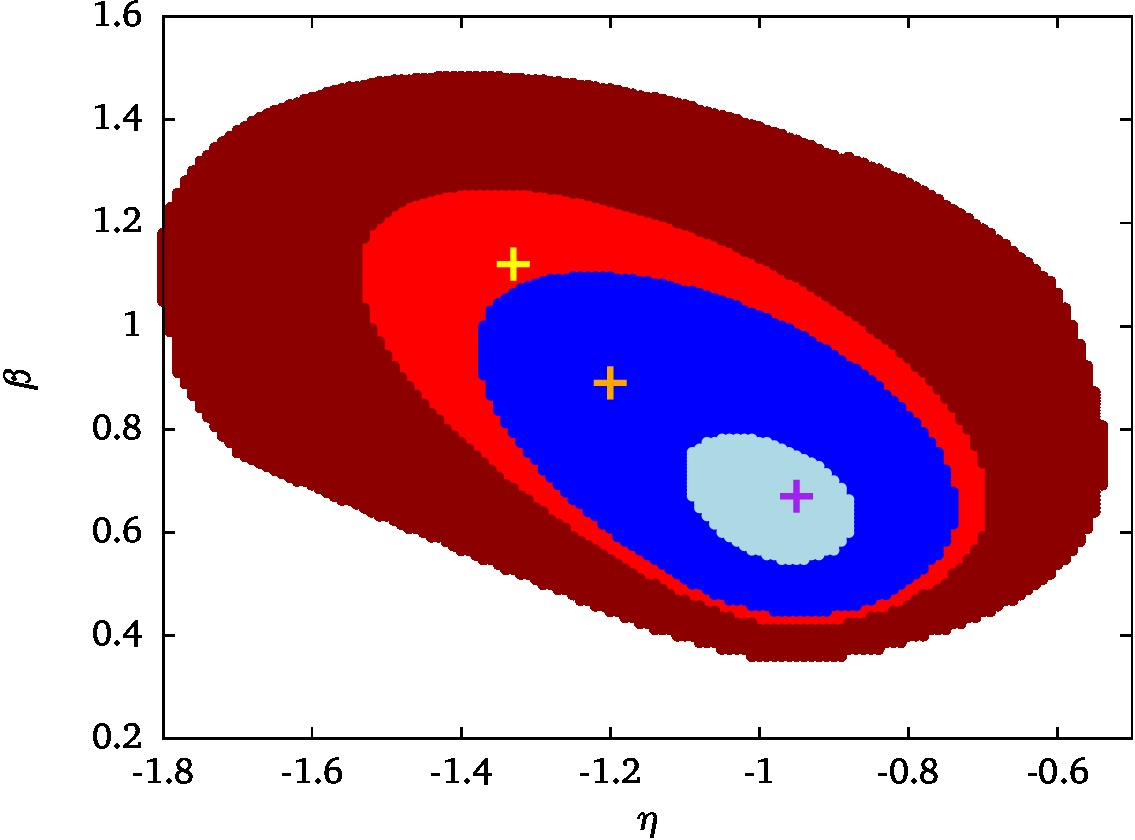}
\end{minipage}
\begin{minipage}{0.48\textwidth}
\includegraphics[width=\textwidth]{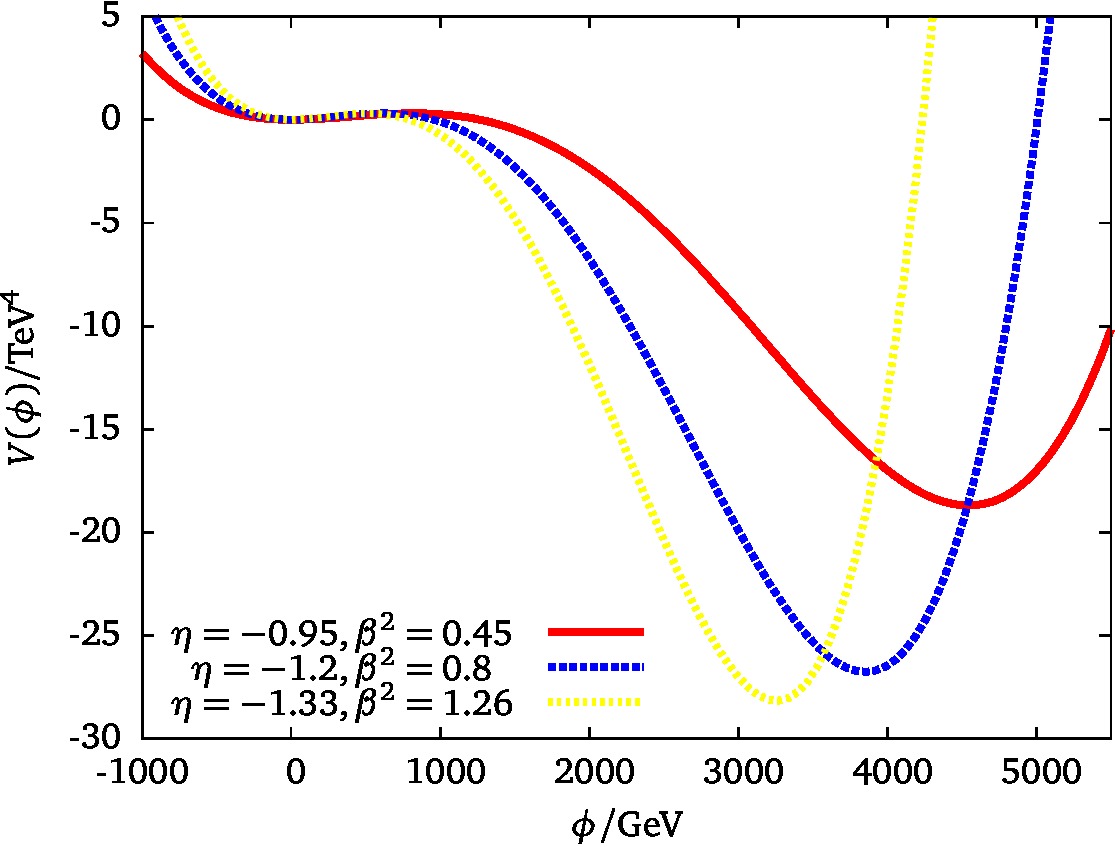}
\end{minipage}
\caption{The field configurations for a certain sample point (\(\mu =
  500\,\GeV\), \(A_\bq = A_\tq = 1500\,\GeV\), \(\tan\beta = 40\) and
  \(M_\text{SUSY} = 1\,\TeV\)), which yields \(m_{h^0} = 126\,\GeV\)
  with \(m_{\tilde g} = 1.5 M_\text{SUSY}\) and \(m_A = 800\,\GeV\) but
  is already excluded by the traditional CCB bound for \(A_\bq\) (the
  \(A_\tq\)-bound is passed) lead to very different conclusions on the
  stability of the desired vacuum on cosmological scales. While the
  shape of the potential is qualitatively very much the same over the
  excluded field space (roughly \(\eta \in [-1.8, 0.55]\) and \(\beta
  \in [0.13, 2.2]\) and the non-standard \vev{} varying within maybe a
  \(1000, \dotsc, 2000\,\GeV\) range, the bounce action (shown in
  contours on the left panel) indicates cosmologically stable and
  long-lived (blue: \(B > 400\), light blue: \(B > 1000\)) configuration
  as well as meta-stable and very short-lived (red: \(B < 400\), dark
  red: \(B < 230\), corresponding to a life-time of less then a
  second). The crosses on the left-side plot denote positions of the
  three choices in \(\eta\) and \(\beta\) shown on the right; the yellow
  one corresponds to the yellow line, for the others we have
  \(\text{orange} = \text{blue}\) and \(\text{purple} = \text{red}\).
}\label{fig:bounceample}
\hrule
\end{figure}

\begin{figure}
\begin{minipage}{0.48\textwidth}
\includegraphics[width=\textwidth]{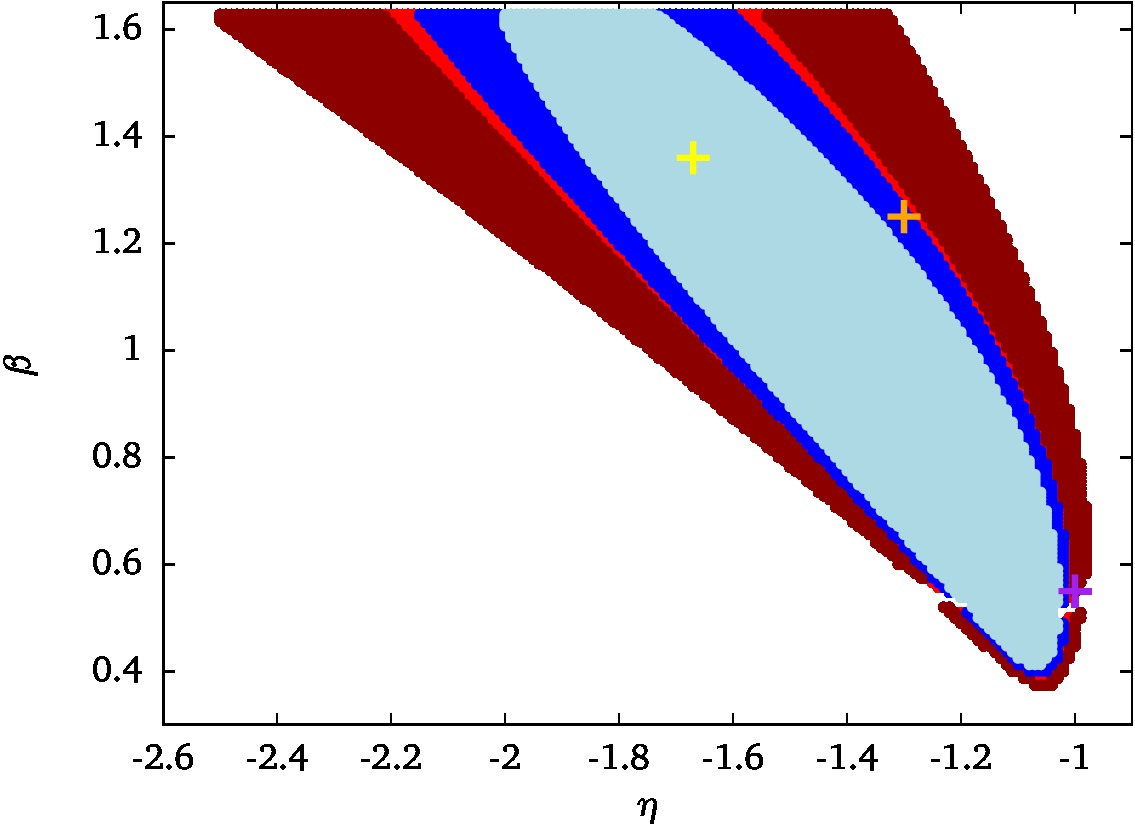}
\end{minipage}
\begin{minipage}{0.48\textwidth}
\includegraphics[width=\textwidth]{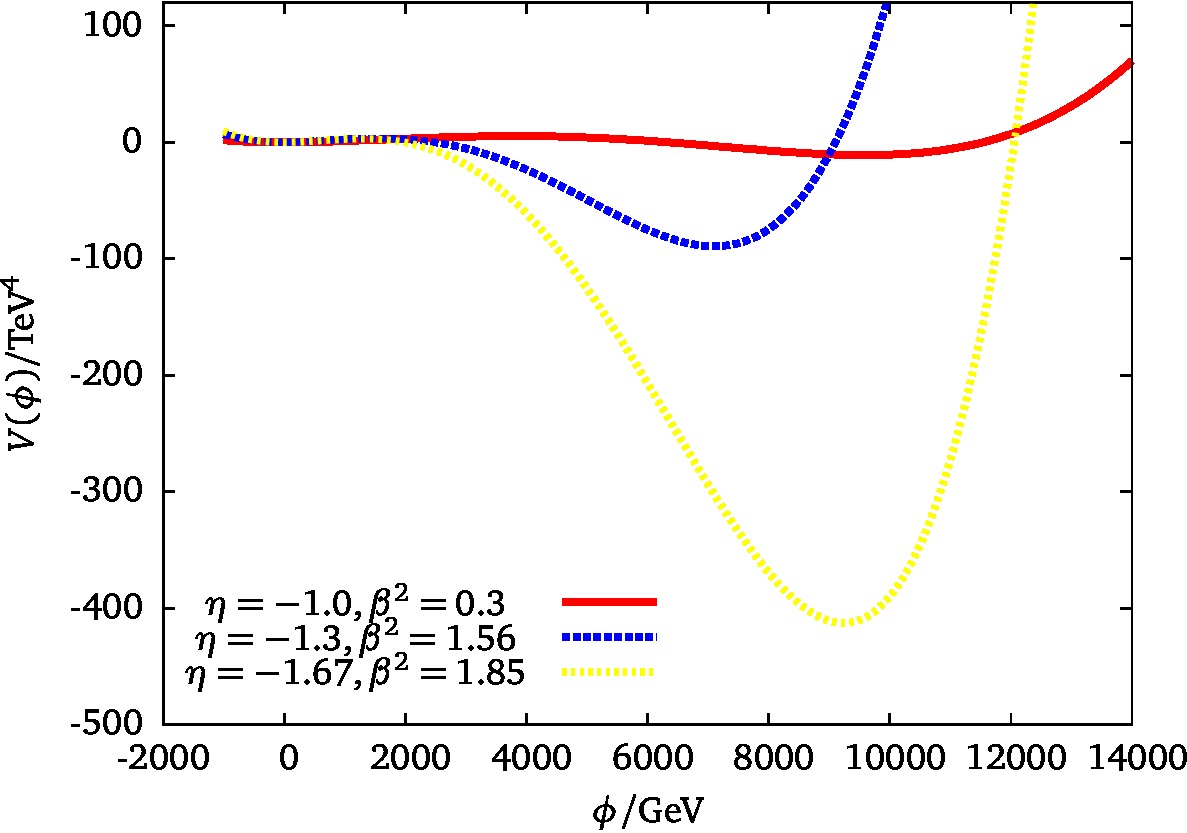}
\end{minipage}
\caption{The same as for Fig.~\ref{fig:bounceample-severe} but a point
  which has a long-lived desired vacuum w.r.t.\ the true vacuum and a
  too light Higgs of \(m_{h^0} = 113\,\GeV\). We have \(A_\tq = A_\bq =
  \mu = 500\,\GeV\), \(\tan\beta = 40\). All other parameters and color
  coding as in Fig.~\ref{fig:bounceample-severe}. Here, the global
  minimum (indicated by the yellow cross in the light blue area) would
  suggest the desired vacuum to be extremely long-lived. This conclusion
  may be misleading as there are other configuration with a much shorter
  tunneling time.}
\label{fig:bounceample-severe}
\hrule
\end{figure}

\begin{figure}
\begin{minipage}{.32\textwidth}
\includegraphics[width=\textwidth]{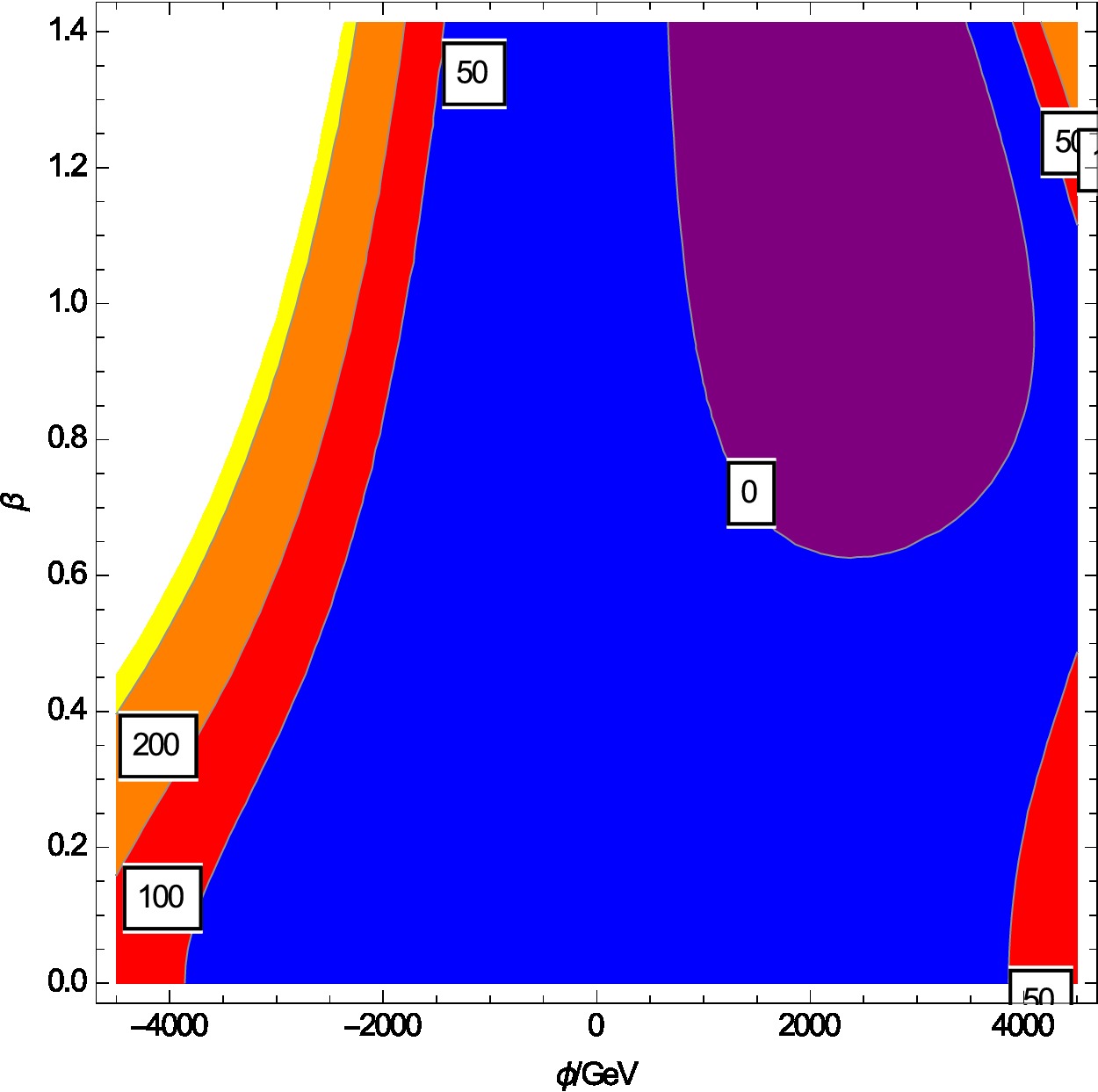}
\end{minipage}
\hfill
\begin{minipage}{.32\textwidth}
\includegraphics[width=\textwidth]{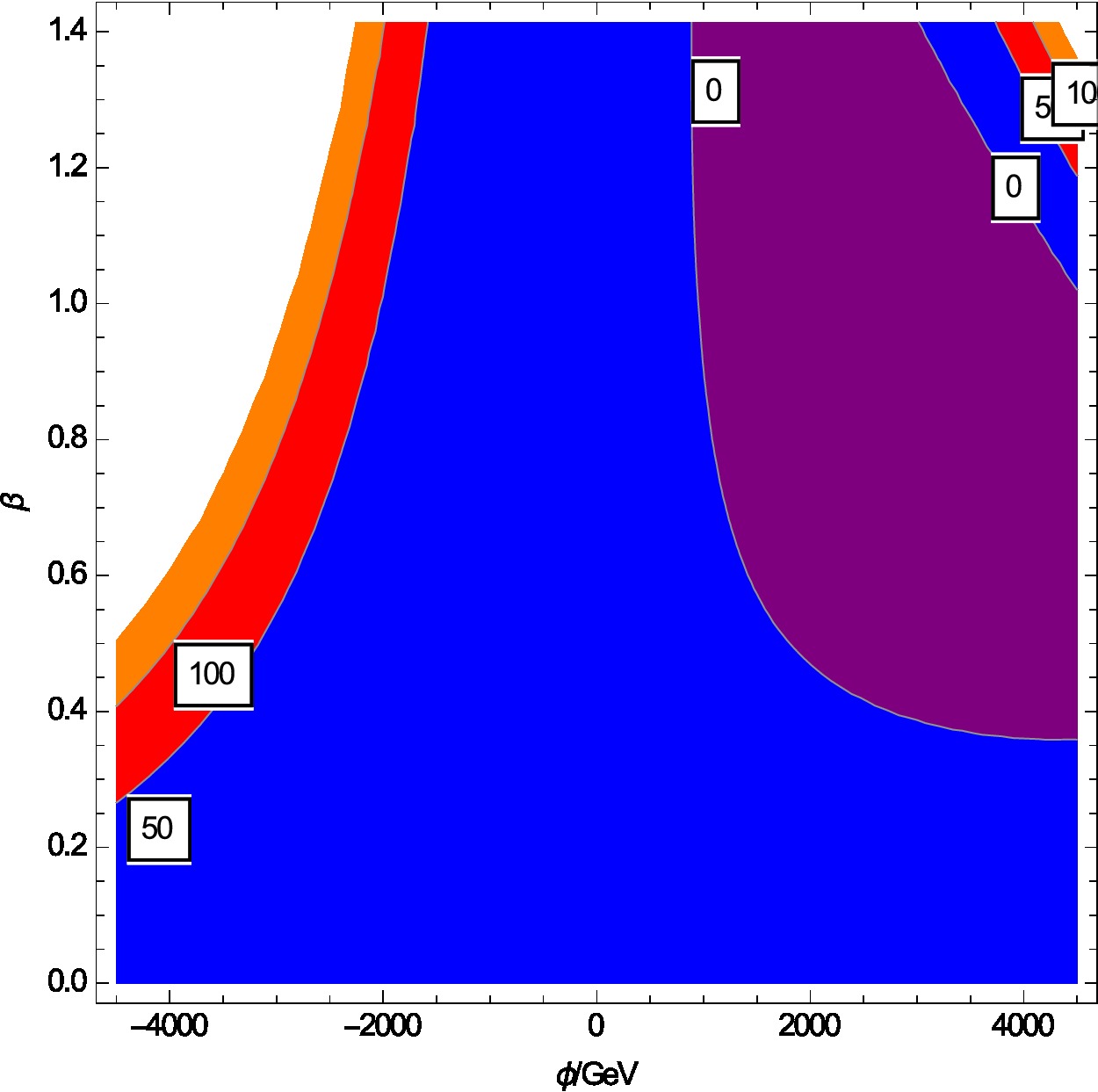}
\end{minipage}
\hfill
\begin{minipage}{.32\textwidth}
\includegraphics[width=\textwidth]{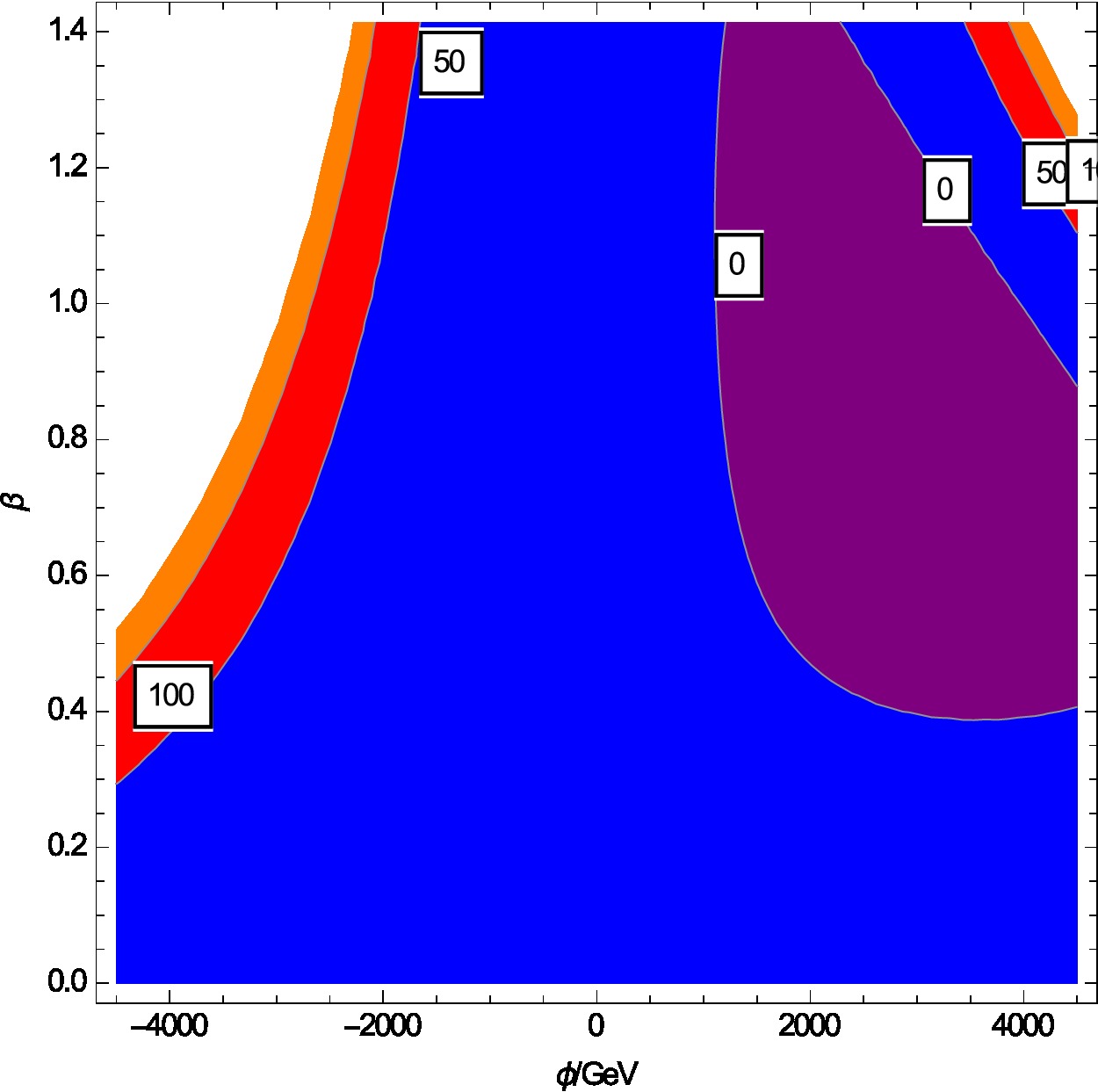}
\end{minipage}\\[.2em]
\begin{minipage}{.32\textwidth}
\includegraphics[width=\textwidth]{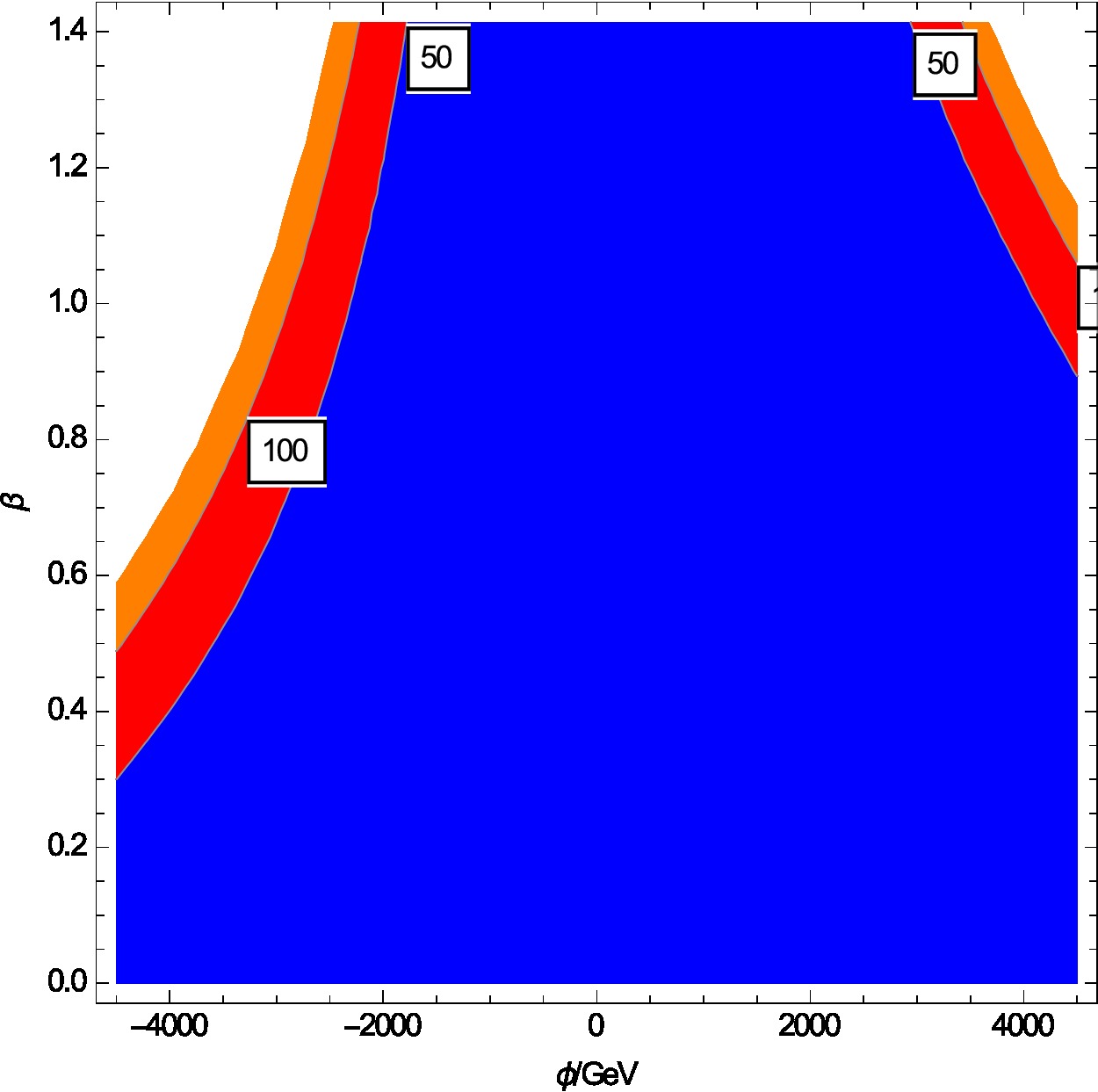}
\end{minipage}
\hfill
\begin{minipage}{.32\textwidth}
\includegraphics[width=\textwidth]{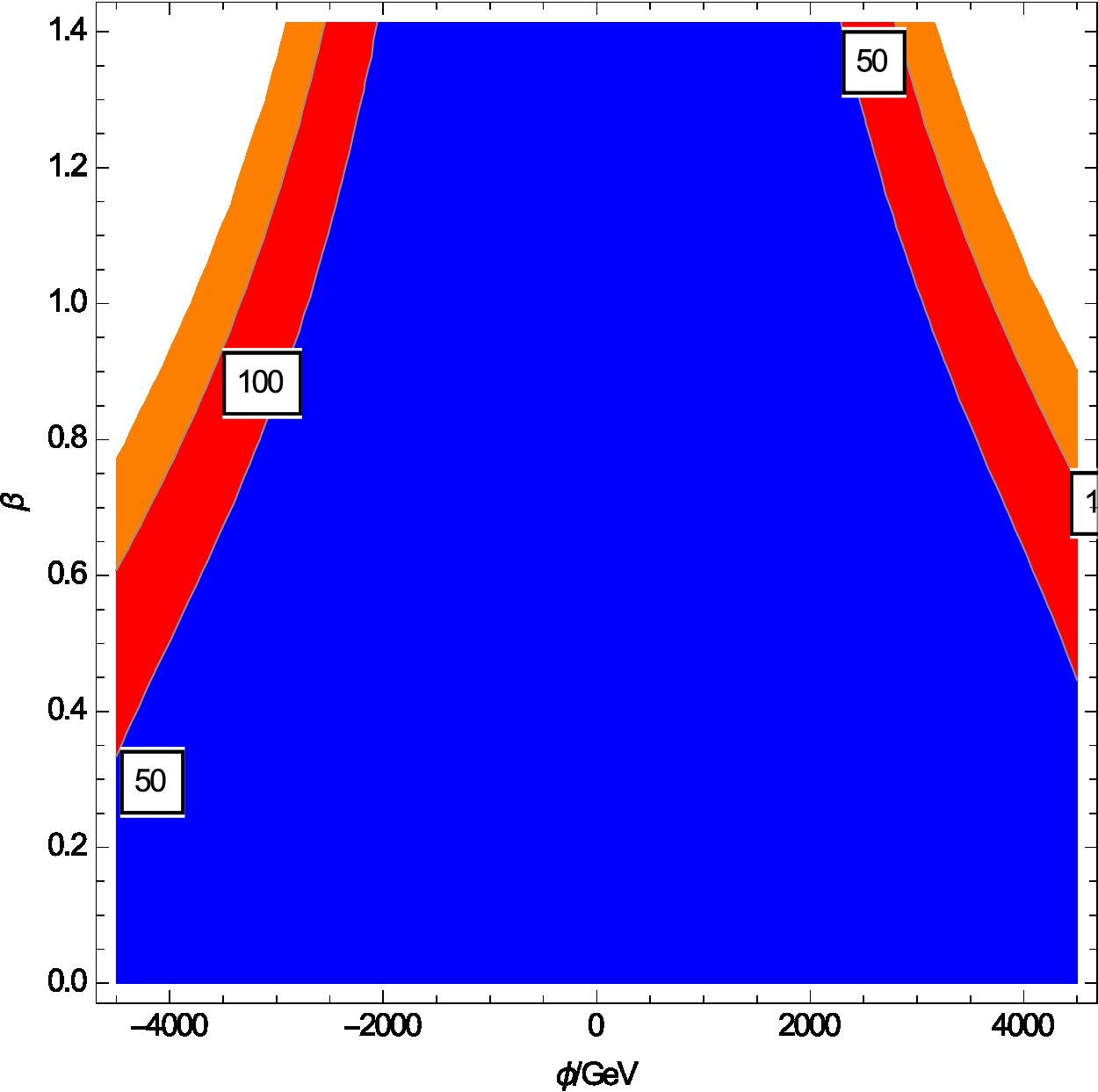}
\end{minipage}
\hfill
\begin{minipage}{.32\textwidth}
\includegraphics[width=\textwidth]{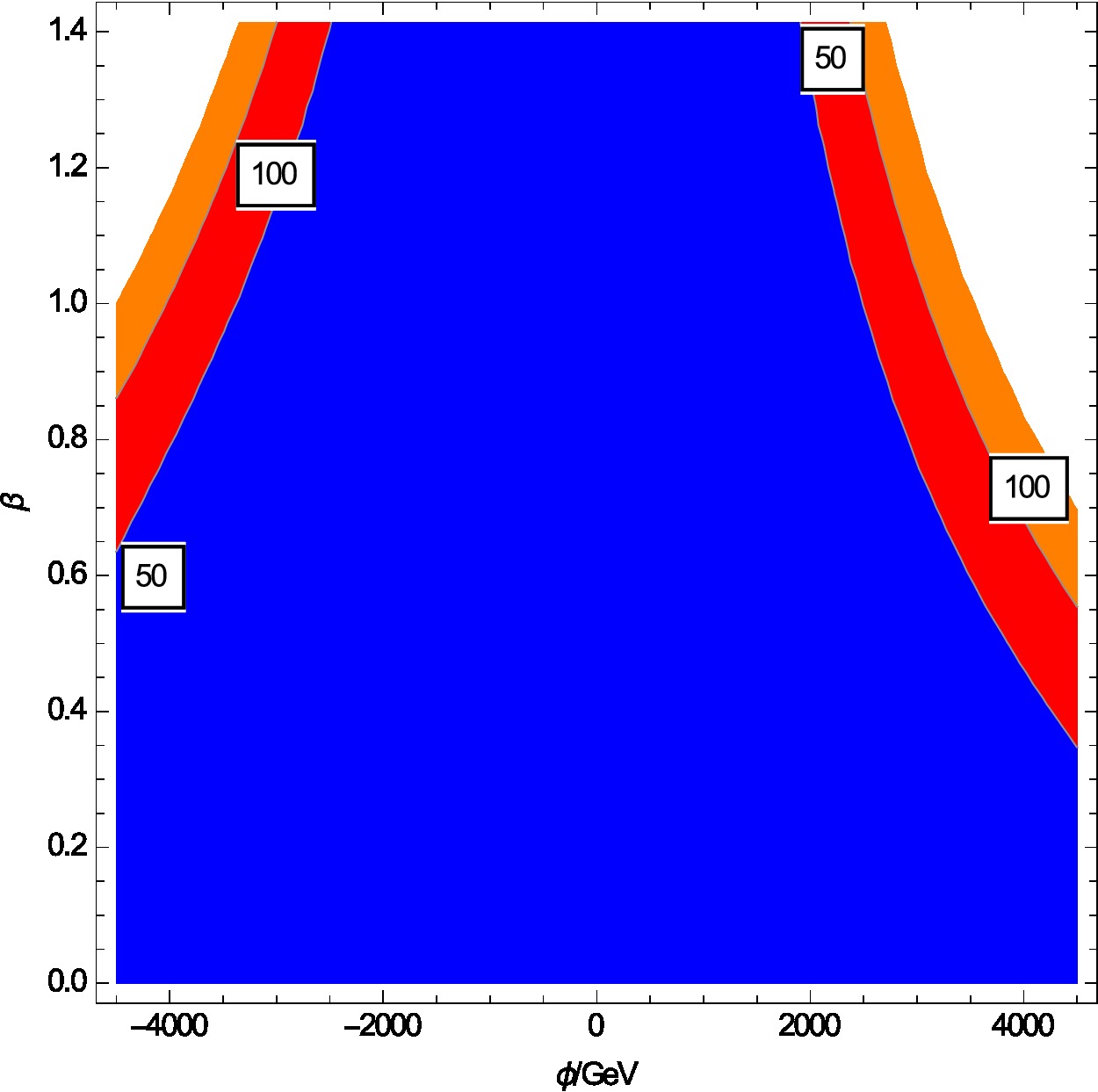}
\end{minipage}\\[.2em]
\begin{minipage}{.32\textwidth}
\includegraphics[width=\textwidth]{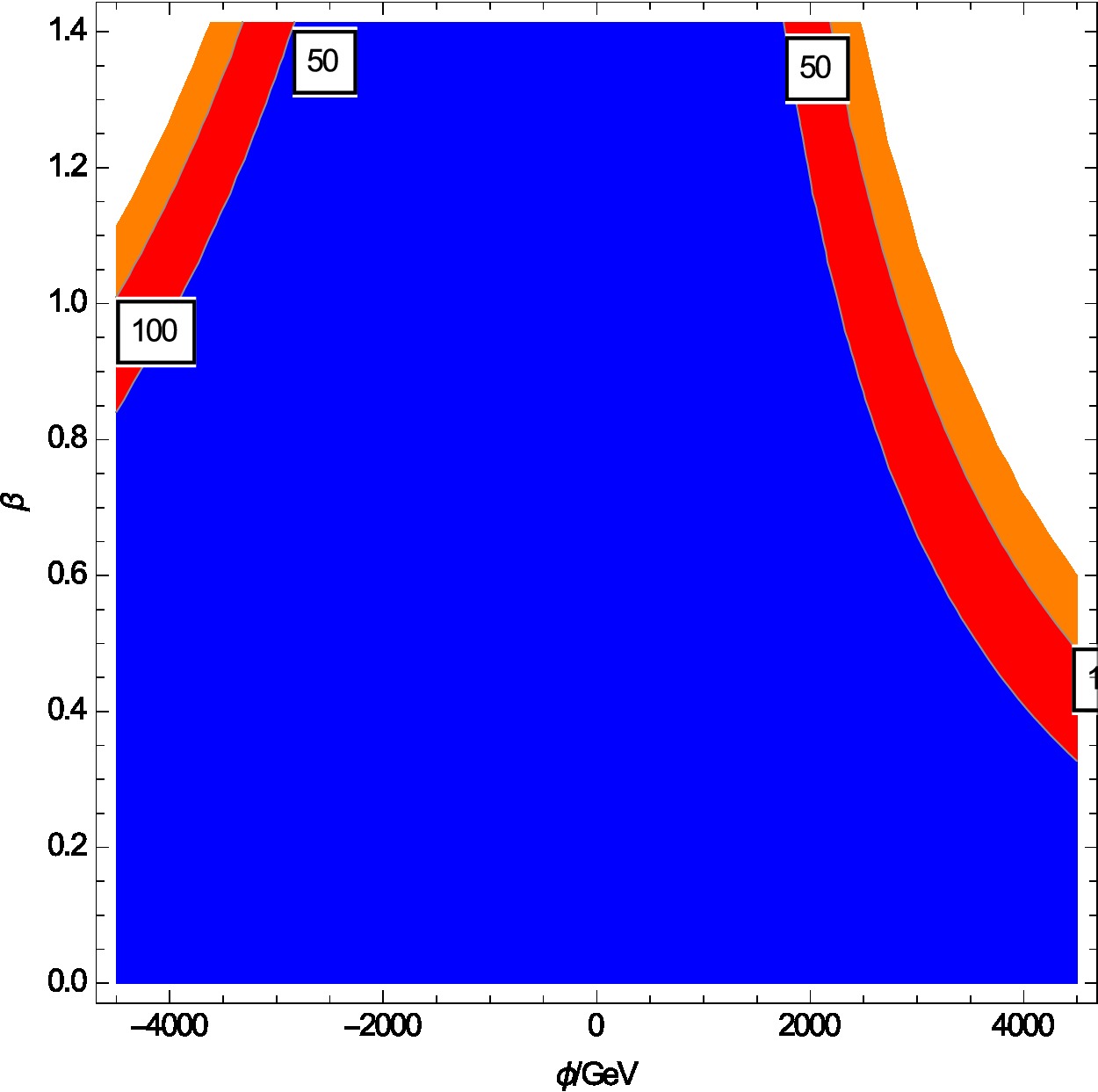}
\end{minipage}
\hfill
\begin{minipage}{.32\textwidth}
\includegraphics[width=\textwidth]{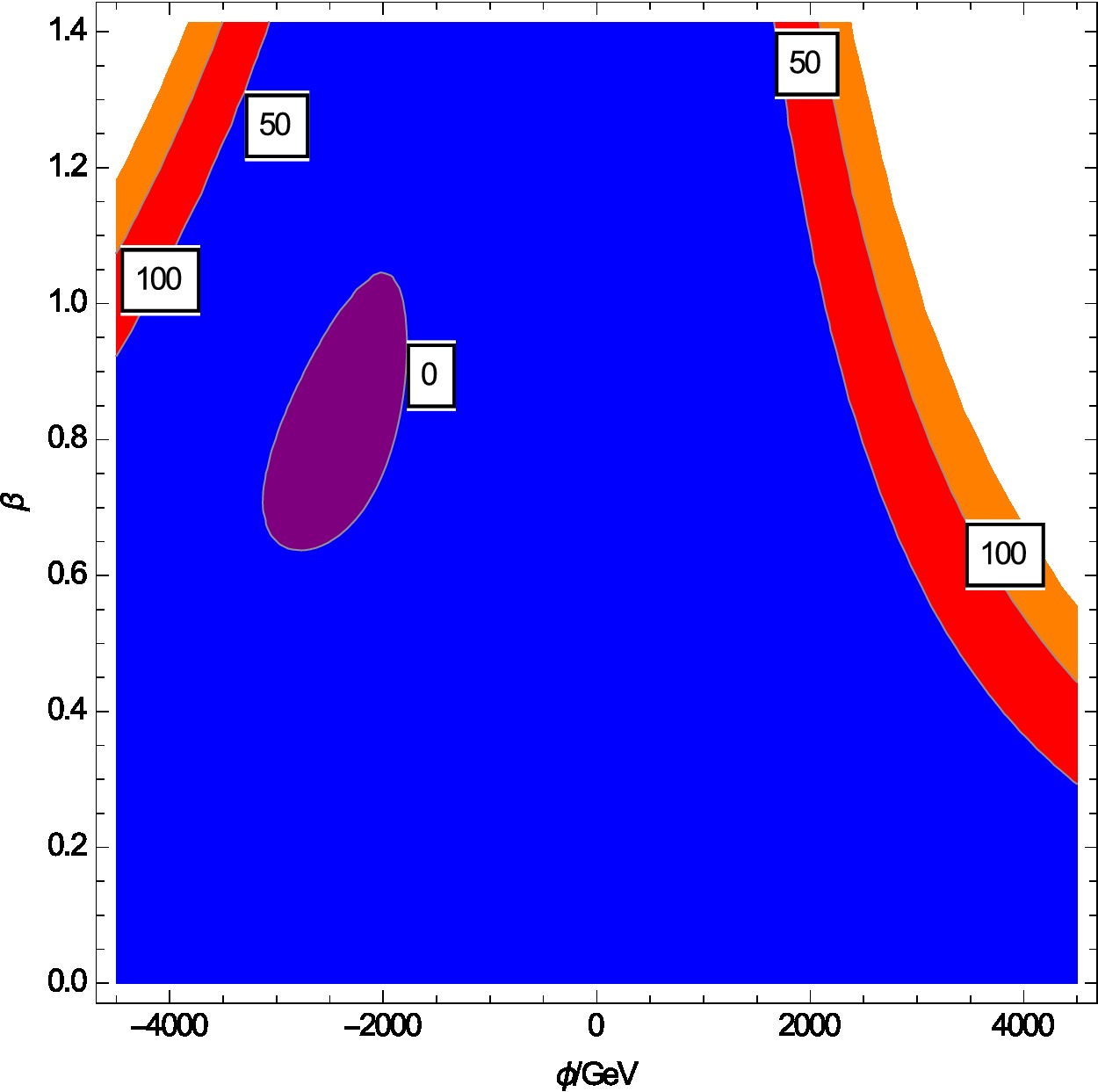}
\end{minipage}
\hfill
\begin{minipage}{.32\textwidth}
\includegraphics[width=\textwidth]{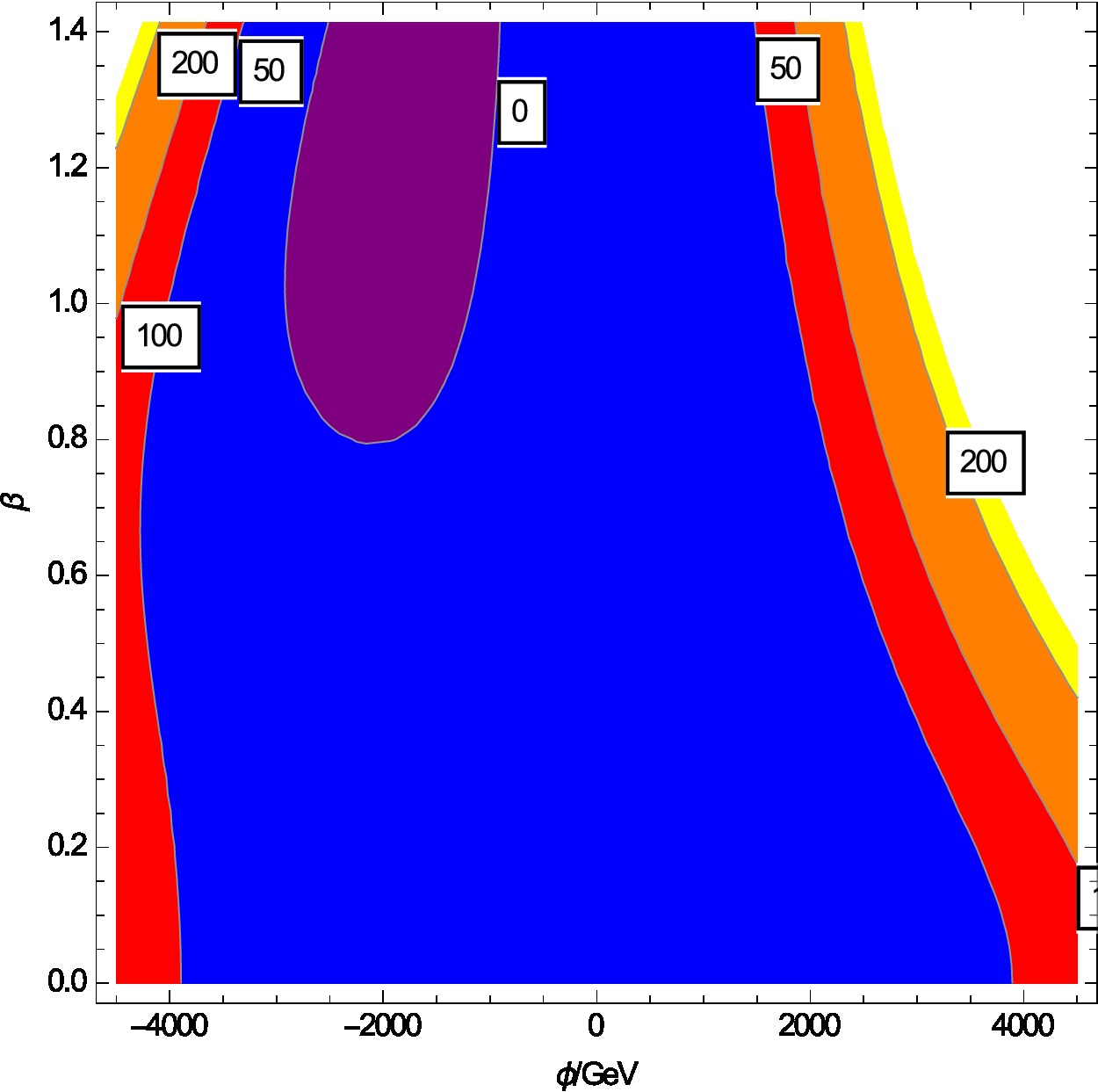}
\end{minipage}\\
\caption{A tomographicly sliced view of the CCB potential in the
  \(\phi = h_\uq\) and \(\tilde b = \beta \phi\) direction for \(\eta =
  {-1.5, -1.0, -0.8, -0.5, 0, 0.5, 0.8, 1.0, 1.5}\) (reading single
  plots from left to right and top to bottom), where \(h_\dq = \eta
  \phi\). The numbers at the contour lines represent the scaled potential
  value \(V(\phi, \eta, \beta) / \TeV^4\) to enhance
  readability. Negative regions within the \(0\)-contour indicate the
  existence of a non-standard true vacuum.}
\label{fig:tomo_eta}
\hrule
\end{figure}

\begin{figure}
\begin{minipage}{.32\textwidth}
\includegraphics[width=\textwidth]{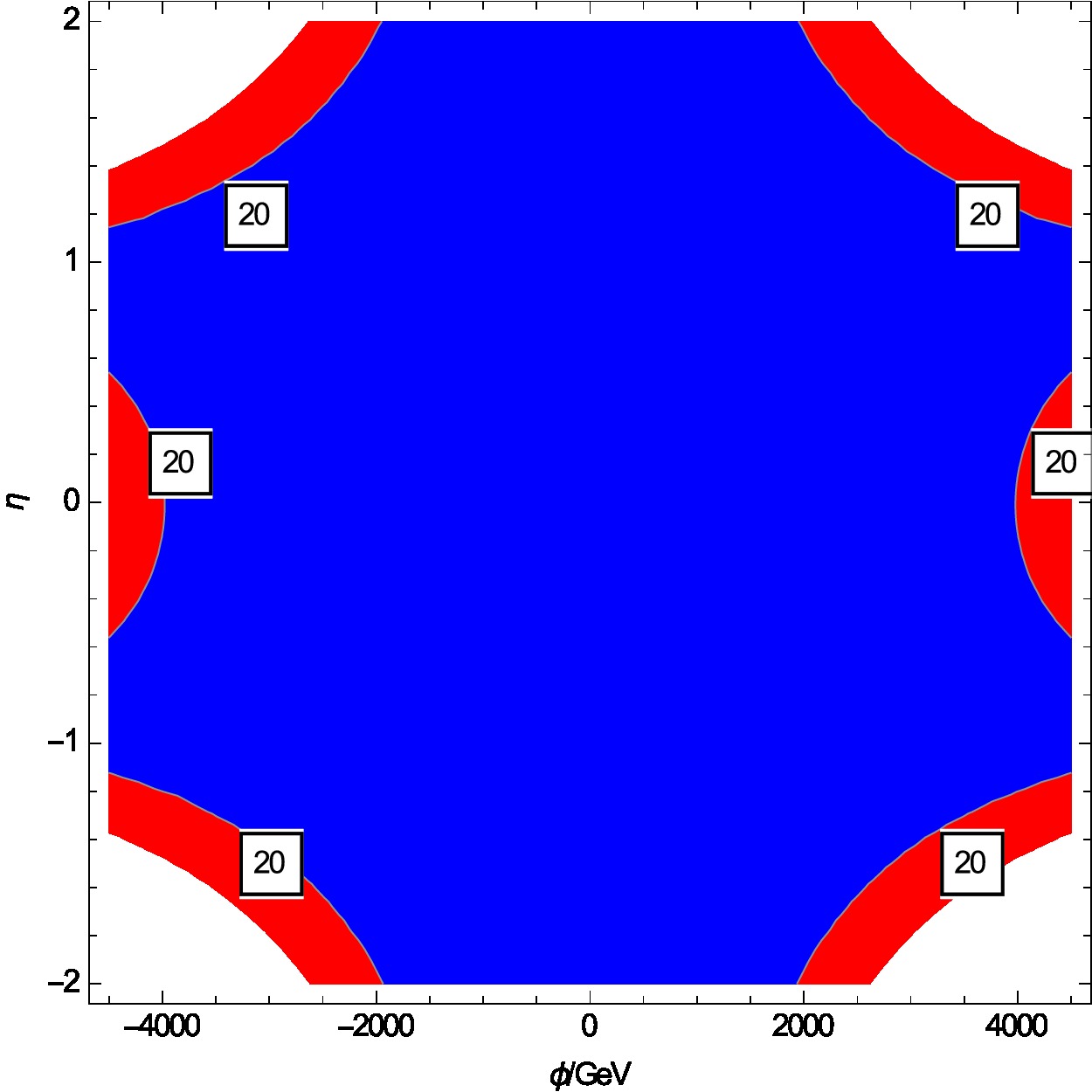}
\end{minipage}
\hfill
\begin{minipage}{.32\textwidth}
\includegraphics[width=\textwidth]{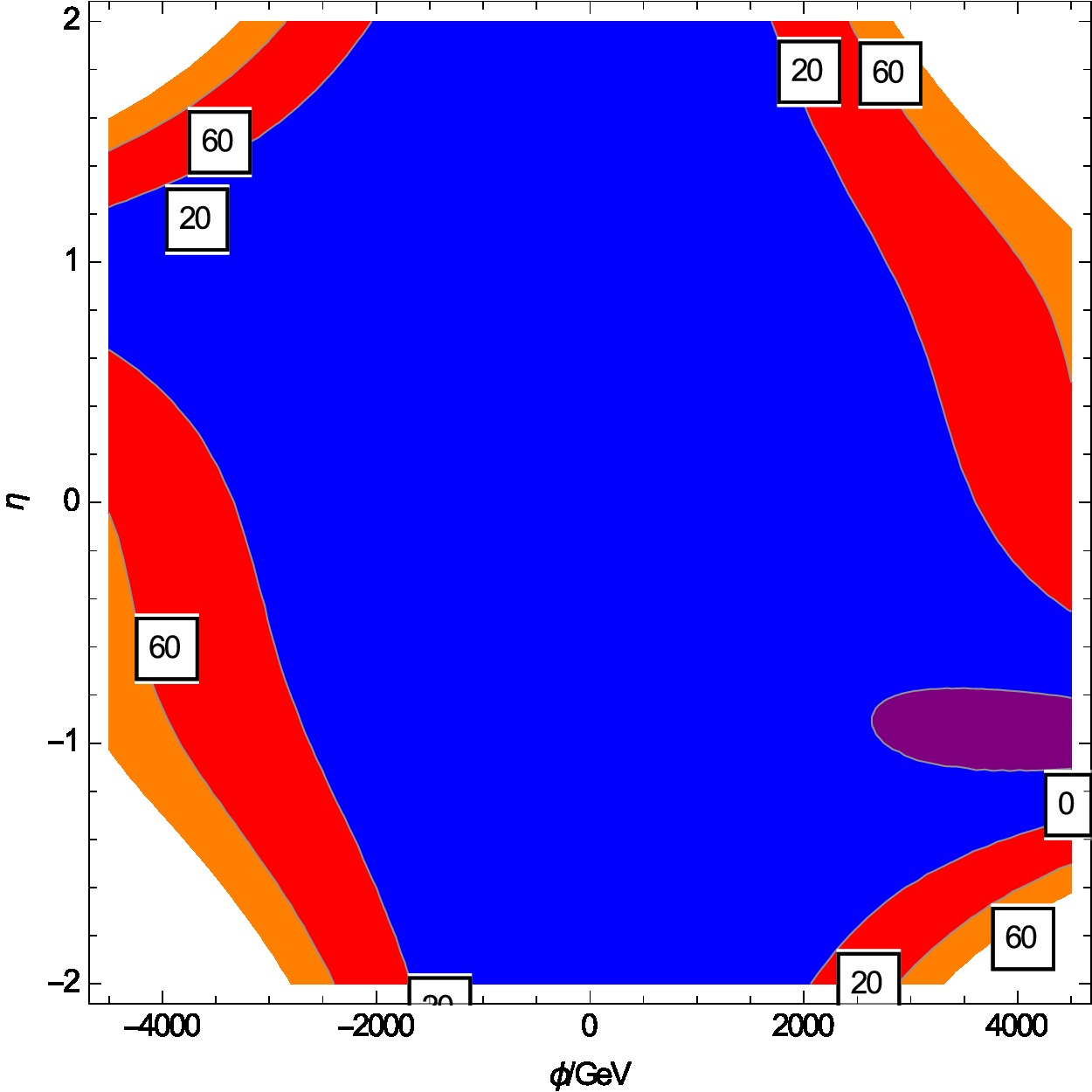}
\end{minipage}
\hfill
\begin{minipage}{.32\textwidth}
\includegraphics[width=\textwidth]{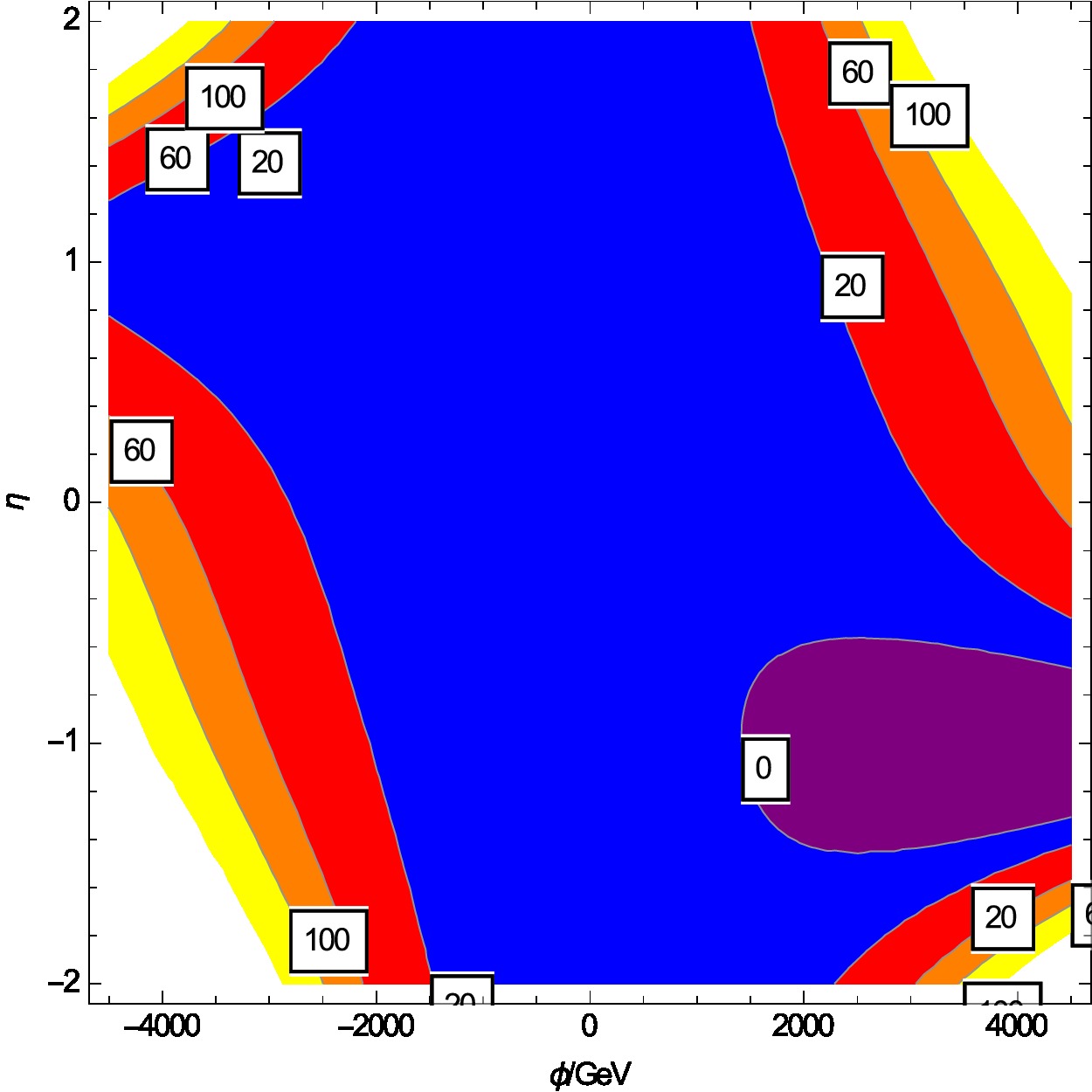}
\end{minipage}\\[.2em]
\begin{minipage}{.32\textwidth}
\includegraphics[width=\textwidth]{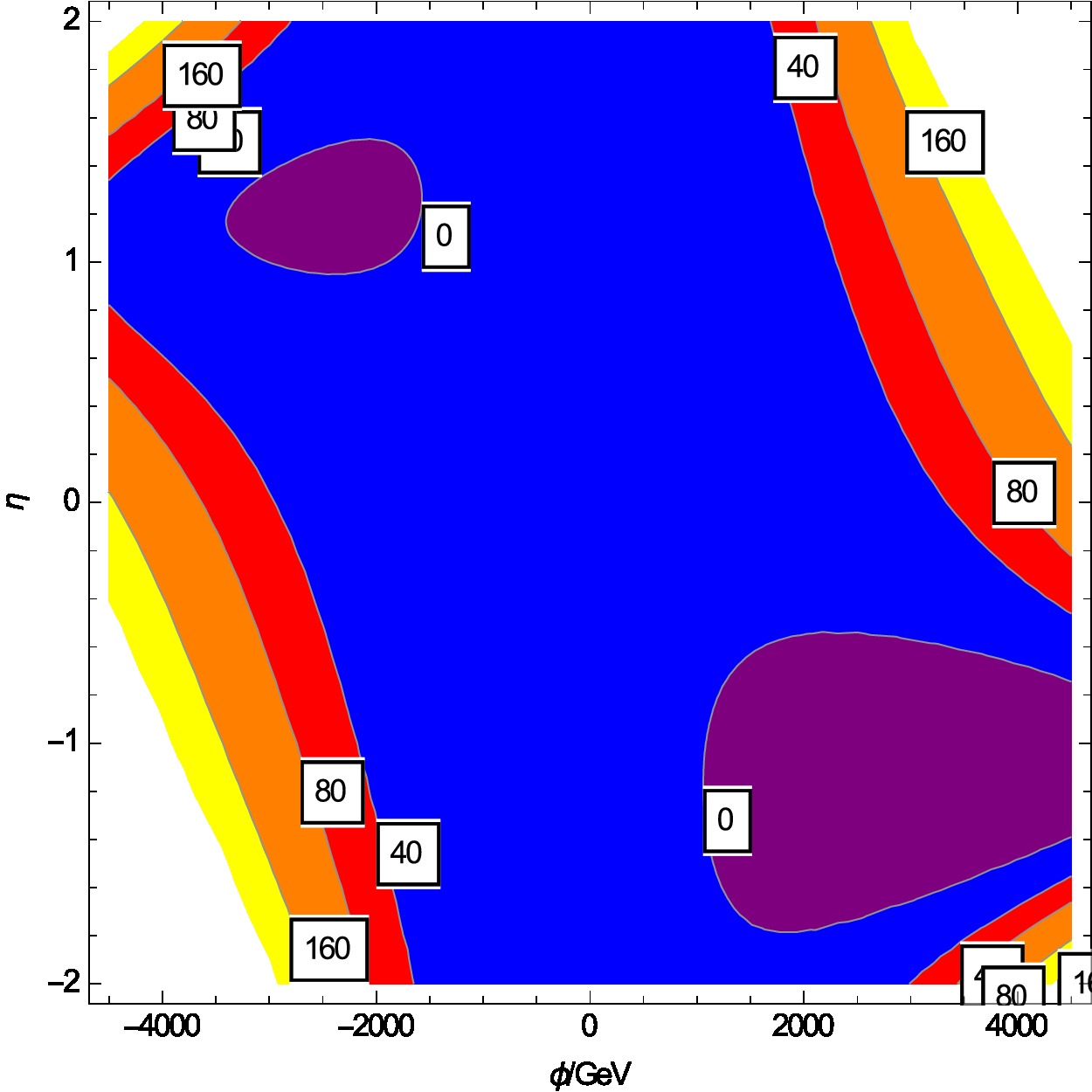}
\end{minipage}
\hfill
\begin{minipage}{.32\textwidth}
\includegraphics[width=\textwidth]{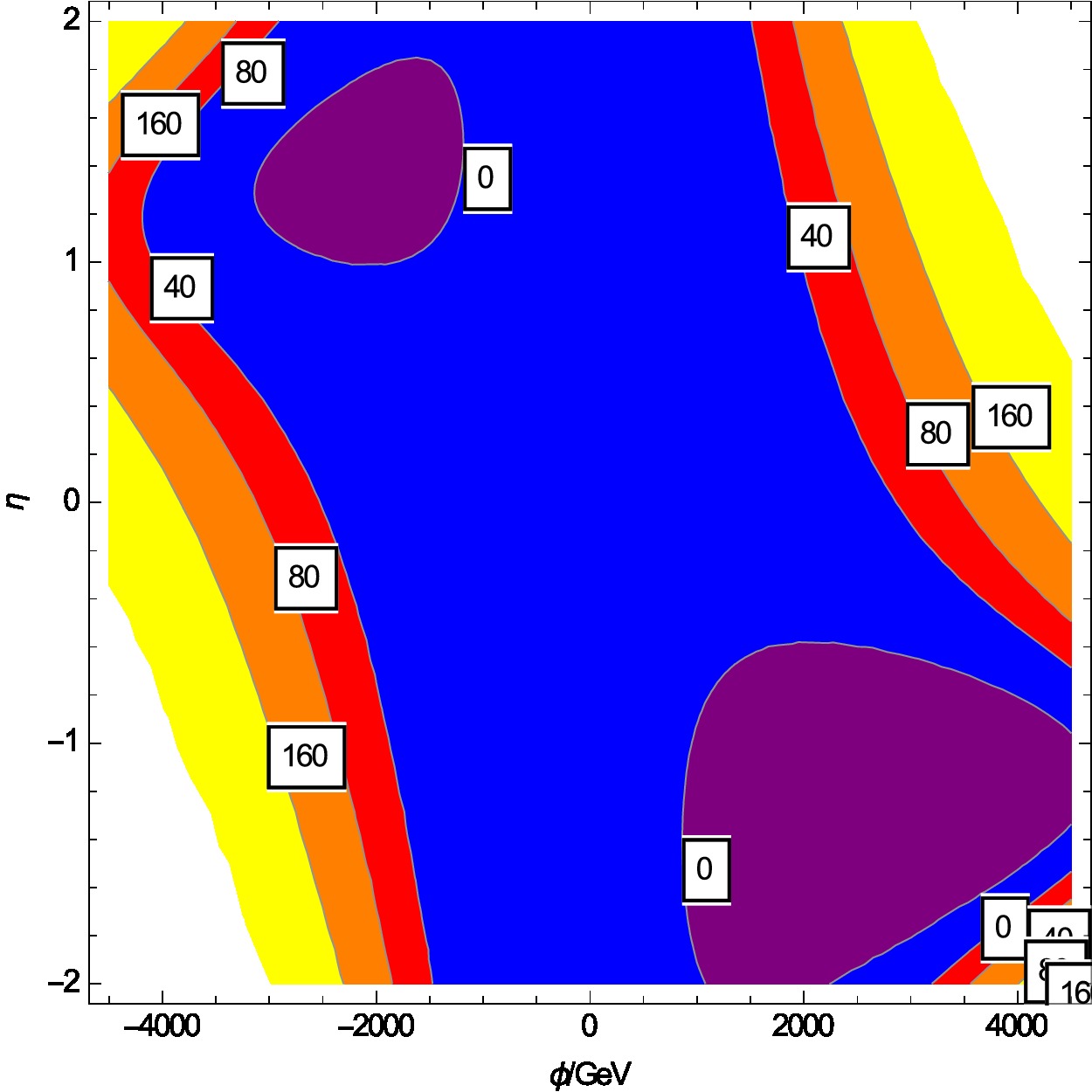}
\end{minipage}
\hfill
\begin{minipage}{.32\textwidth}
\includegraphics[width=\textwidth]{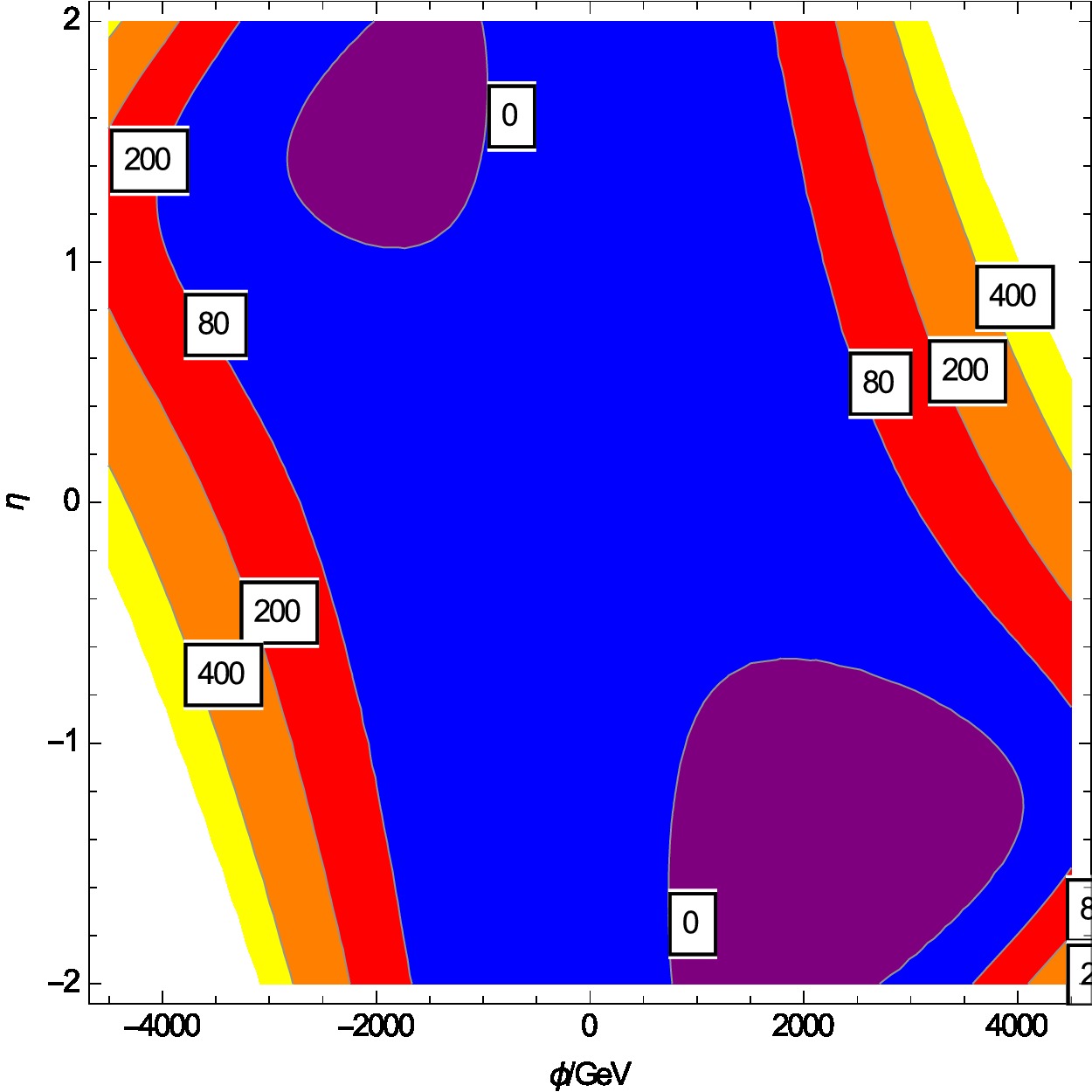}
\end{minipage}\\[.2em]
\begin{minipage}{.32\textwidth}
\includegraphics[width=\textwidth]{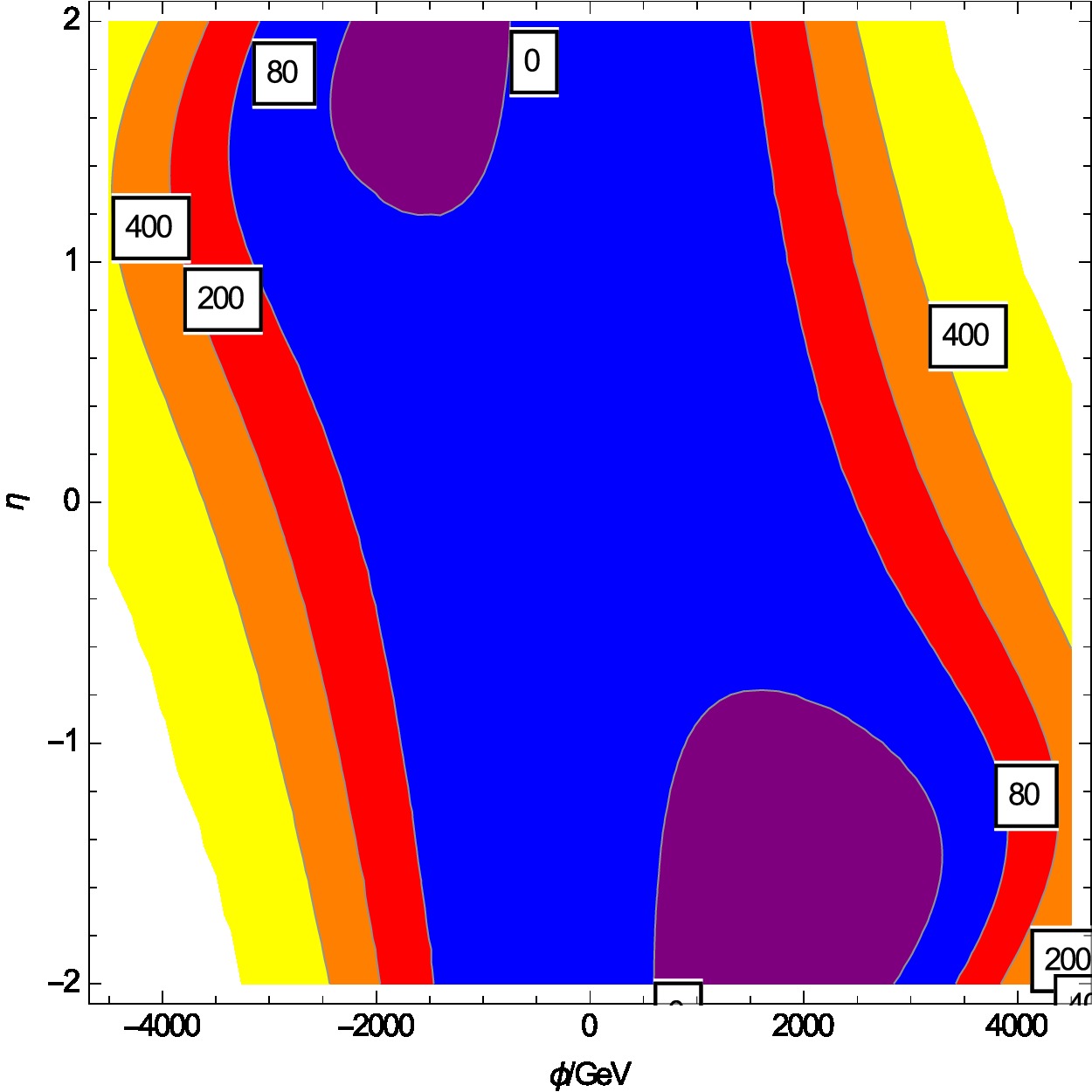}
\end{minipage}
\hfill
\begin{minipage}{.32\textwidth}
\includegraphics[width=\textwidth]{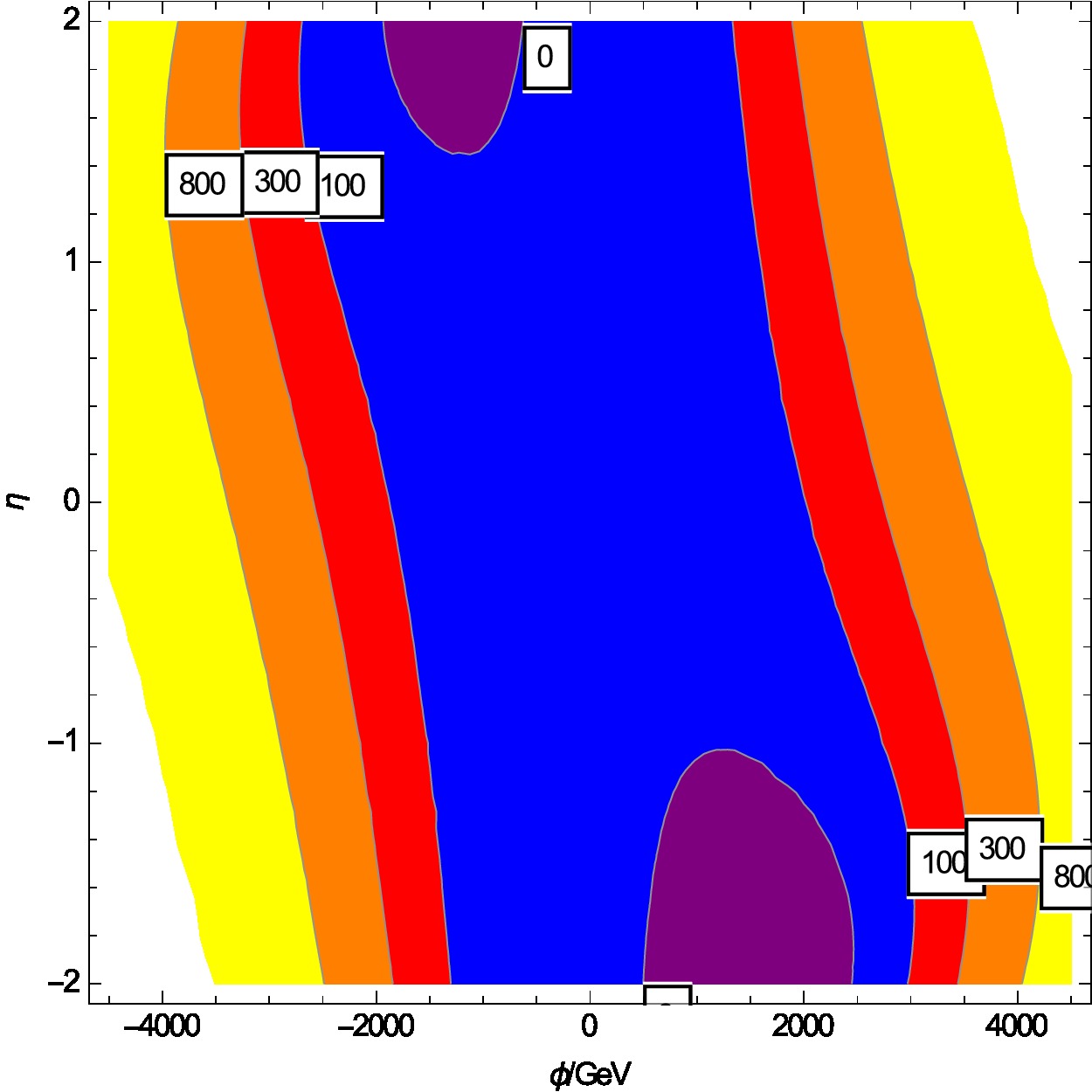}
\end{minipage}
\hfill
\begin{minipage}{.32\textwidth}
\includegraphics[width=\textwidth]{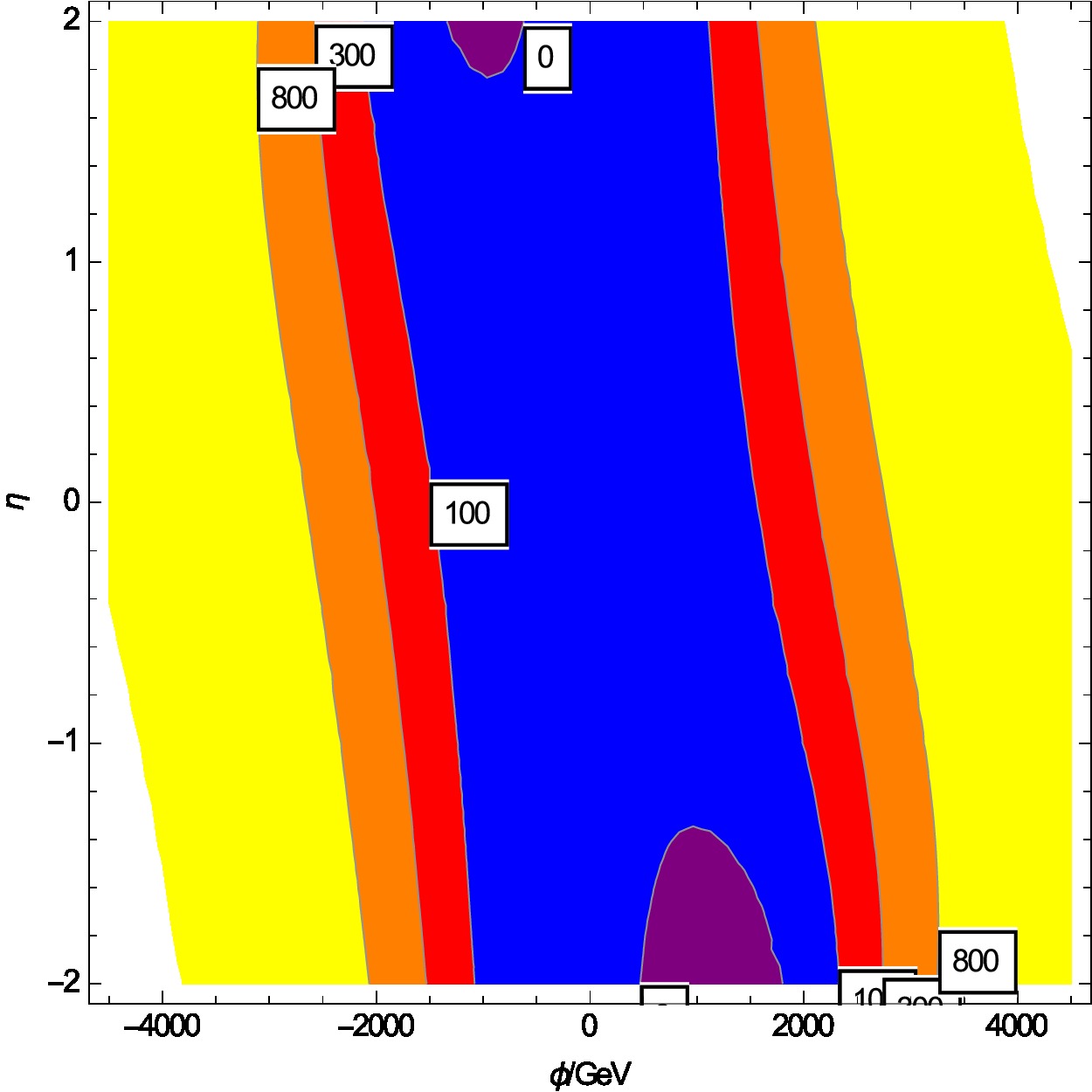}
\end{minipage}\\[.2em]
\caption{A similar tomographic view as in Fig.~\ref{fig:tomo_eta} in
  the perpendicular direction with \(\eta = h_\dq / \phi\) on the
  vertical axis and scanning \(\beta = \tilde b / \phi\) for \(\beta =
  {0, 0.4, 0.6, 0.8, 1.0, 1.2, 1.5, 2.0, 2.6}\) from left to right and
  top to bottom with \(\phi = h_\uq\). The existence of up to two
  non-standard vacua reflects the actually broken reflection symmetry,
  \(\eta \to -\eta\) and \(\phi \to -\phi\), as can be seen that there
  no such exact symmetry.}
\label{fig:tomo_beta}
\hrule
\end{figure}

To check for metastability, one may be tempted to define the field
configuration and the specific ray that shows the deepest non-standard
vacuum as the ideal or optimal one. However, as the new \vev{} appears
at say \(\mathcal{O}(10\,\TeV)\) and the barrier in between gets
sufficiently high, say \(\mathcal{O}(\text{few}~\TeV^4)\), \(B\) is
\(\gg 400\) in that specific direction as for the point in
Fig.~\ref{fig:bounceample-severe}. However, there are other directions
via which the global minimum can be accessed with a much smaller
tunneling time. For the sample point of Fig.~\ref{fig:bounceample} from
above, we show a tomographic view of the scalar potential in the
\(\tilde b\)-\(h_\uq\) plane for increasing \(\eta = h_\dq / h_\uq\) in
Fig.~\ref{fig:tomo_eta} and the same potential sliced differently for
increasing \(\beta = \tilde b / h_\uq\) in the \(h_\dq\)-\(h_\uq\) plane
in Fig.~\ref{fig:tomo_beta}. This is to illustrate that there is no
unique choice for some fixed values of \(\eta\) and \(\beta\) that
exclusively show a non-standard vacuum. There are wide regions in field
space and all paths should be treated equal to estimate the tunneling
rate. The ``optimal'' direction for the determination of the bounds on
the potential parameters (masses, trilinear and quadrilinear couplings)
should be rather given by the shortest tunneling time. As recommendation
how to deal with \emph{any} CCB exclusion, we declare each point that
fails the condition
\[
\mathcal{A}(\eta, \alpha, \beta)^2 < 4 \lambda(\eta, \alpha, \beta)
M^2 (\eta, \alpha, \beta)
\]
for \emph{any} specific combination of \(\eta\), \(\alpha\) and
\(\beta\) as clearly unstable. An easy (but maybe CPU intensive) way to
check this is to scan over a reasonable range, \eg{} \(\eta \in [-3,
3]\) and \(\alpha, \beta \in [0, 2]\); with a binning of \(0.1\) this
procedure should find CCB configurations (since the field space regions
are quite extended, even coarser binnings should lead to a trustable
result).

%%%%%%%%%% FIGURE %%%%%%%%%%
%%%%%%%%%% % eta tomography %%%
%%%%%%%%%% FIGURE %%%%%%%%%%
%%%%%%%%%%%%%

%%%%%%%%%% FIGURE %%%%%%%%%%
%%%%%%%%%% % beta tomography %%%
%%%%%%%%%% FIGURE %%%%%%%%%%
%%%%%%%%%%%%%

\section{Anatomy of Charge and Color Breaking}
\label{sec:anat}

The main stability condition is given by the unequation
\begin{equation}\label{eq:generic-bound}
M^2(\eta, \alpha, \beta) > \frac{A(\eta, \alpha,
  \beta)^2}{4\lambda(\eta, \alpha, \beta)},
\end{equation}
depending on the field misalignments, as well as on all relevant model
parameters. We distinguish between parameters in configuration space
that independently of the model parameters can lead to instable
configurations, such as \(\alpha\), \(\beta\) and \(\eta\)---and the
model parameters that change the shape of the scalar potential as whole
object (such as the soft masses, the trilinear couplings \(A_\tq\) and
\(A_\bq\) as well as the \(\mu\)-parameter). Furthermore, we request the
parameters of the one-loop Higgs potential (\ie the genuine type-II 2HDM
of the MSSM) to allow for spontaneous electroweak breaking with the
correct \vev{}s. The ratio of the \emph{standard} \(v_\uq/v_\dq\) is
what we call \(\tan\beta\), obeying \(v_\dq^2 + v_\uq^2 = v^2 =
(246\;\GeV)^2\). Note that finally, the ``true'' \(\tan\beta\) will be
given by \(1/\eta\). We neither keep \(\tan\beta\) fixed in a sense that
the true vacuum has to respect this relation, nor do we infer that from
an original \(\tan\beta > 1\) the ratio \(\langle h_\uq \rangle /
\langle h_\dq \rangle\) has to have the same property. What we actually
find is that for most configurations the true vacuum seems to have
\(\langle h_\dq \rangle > \langle h_\uq \rangle\) and the CCB \vev{}
typically shows \(\langle \tilde q \rangle \gtrsim 0.7 \langle
h_\uq\rangle\). Unfortunately, for the bounds deduced numerically by
attempting to find the global minimum of parameter points
violating~\eqref{eq:generic-bound}, no expression of the scaling
parameters \(\alpha\), \(\beta\) and \(\eta\) can be found in terms of
the relevant potential parameters although they crucially depend on the
MSSM parameter point. The ideal solution would be an exclusion of the
form~\eqref{eq:generic-bound} with \(\eta\), \(\alpha\) and \(\beta\)
given in terms of the model parameters. Similar attempts have been
achieved by~\cite{Casas:1995pd} where one trilinear operator at a time
was considered only. For four operators this appears to be impossible.

For the numerical analysis in the following, we consider a very
phenomenological version of the MSSM with all SUSY breaking parameters
defined at the low (SUSY) scale without referring to any high-scale
unified scenario. As the developing CCB minima typically also show up
around the same \emph{low} scale, we ignore any effects from the
renormalization group as the corresponding logarithms are small and only
have a mild impact on the shape of the potential (see
\eg~\cite{Kusenko:1996jn} and the reference therein
to~\cite{Bordner:1995fh}). For the qualitative discussion this point is
irrelevant anyway. Quantitatively, if desired, parameters at the relevant
scale can be employed as input values for the analytical bounds.

We determine the soft SUSY breaking Higgs masses \(m_{H_\uq}^2\) and
\(m_{H_\dq}^2\) requiring electroweak symmetry breaking via the
conditions \(\partial V _1/ \partial h_\uq \big|_{h_\uq = v_\uq, h_\dq =
  v_\dq} = 0\) and \(\partial V _1/ \partial h_\dq \big|_{h_\uq = v_\uq,
  h_\dq = v_\dq} = 0\) with the one-loop Higgs potential
\(V_1\)~\cite{Bobrowski:2014dla}. The bilinear soft breaking term is
related to the pseudoscalar mass at tree-level via \(B_\mu = m_A^2
\sin\beta\cos\beta\). Our free parameters are the ratio of the two Higgs
\vev{}s at tree-level, \(\tan\beta = v_\uq / v_\dq\), the soft squark
masses, which we for simplicity set to \({\tilde m}_Q^2 = {\tilde m}_t^2
= {\tilde m}_b^2 = M_\text{SUSY}^2\), and the superpotential parameter
\(\mu\) as well as the soft breaking trilinear Higgs--squark couplings
\(A_\tq\) and \(A_\bq\). The gaugino masses \(M_1, M_2\) and \(M_3\)
enter only indirectly and play a less crucial role, where the gluino
mass \(M_3\) can be more important for the threshold effects on the
bottom Yukawa coupling and in the two-loop light Higgs mass as provided
by \textsc{FeynHiggs}. If nothing else is stated, we set \(M_1 = M_2 =
M_\text{SUSY} = 1\,\TeV\) and \(M_3 = 1.5 M_\text{SUSY}\).

Including bottom Yukawa effects in the analysis of CCB minima has not
been done to great extend in the literature, as \(y_\bq\) usually is
neglected because of its smallness. However, for large \(\tan\beta\) and
certain other regions in parameter space this cannot be done
anymore. Especially the \(\Delta_b\) resummation for the bottom quark
mass effectively changes the bottom Yukawa coupling dramatically for
such regions. While \(y_\bq\) gets lowered compared to \(m_\bq / v_\dq\)
for large \(\tan\beta\), it grows severely for negative \(\mu\) as can
be seen from the expressions and even runs into a non-perturbative
region (what is a well-known behavior). The reduction at large
\(\tan\beta\) and small but positive \(\mu\) keeps this window open in
the following analysis. We include the dominant contributions to
\(\Delta_b\) from the gluino and the higgsino loop~\cite{Hall:1993gn,
  Carena:1994bv, Pierce:1996zz, Carena:1999py},\footnote{
  \(\displaystyle C_0(x, y, z) = \frac{x^2 y^2 \log\frac{y^2}{x^2} + y^2
    z^2 \log\frac{z^2}{y^2} + x^2 z^2 \log\frac{x^2}{z^2}}{(x^2 -
    y^2)(x^2 - z^2)(y^2 - z^2)}\).}
\begin{subequations}\label{eq:Deltab}
\begin{align}
\Delta_b^\text{gluino} &= \frac{2\alpha_s}{3\pi} \mu M_{\tilde G}
\tan\beta C_0(\tilde m_{\tilde b_1}, \tilde m_{\tilde b_2}, M_{\tilde G}
), \\
\Delta_b^\text{higgsino} &= \frac{y_\tq^2}{16\pi^2} \mu A_\tq
\tan\beta C_0(\tilde m_{\tilde t_1}, \tilde m_{\tilde t_2}, \mu), \label{eq:Deltabhig}
\end{align}
\end{subequations}
and get the corrected bottom Yukawa coupling with \(\Delta_b =
\Delta_b^\text{gluino} + \Delta_b^\text{higgsino}\) as
\begin{equation}\label{eq:botYukres}
y_\bq = \frac{m_\bq}{v_\dq(1+\Delta_b)}.
\end{equation}

Another remark is inevitable on the relevance of the parameters in the
discussion. Usually, when MSSM effects on the Higgs mass are discussed,
the ``stop mixing parameter'' \(X_t = A_\tq/y_\tq - \mu \cot\beta\) is
used to measure the strength of the corrections rescaled with the SUSY
scale, \(X_t / M_\text{SUSY}\). In the maximal mixing scenario, this
parameter is set to a large value, \(X_t = \sqrt{6} M_\text{SUSY}\) what
appears to be in trouble with the exclusions presented here. Although it
would be desired to directly impose constraints on \(X_t\), we are
unable to do so because the two independent Higgs \vev{}s \(v_\dq\) and
\(v_\uq\) have to be treated as dynamical variables and \(\tan\beta\)
cannot be kept fixed. Doing so would lead to wrong conclusions. However,
we can translate the final exclusions we found to an effective exclusion
on \(X_t\) where \(\tan\beta = v_\uq / v_\dq\) gives the ratio of the
two Higgs \vev{}s in the desired electroweak state. This is especially
important in connection to the importance of \(X_t\) for the light Higgs
mass.

\begin{figure}
\begin{minipage}{0.48\textwidth}
\includegraphics[width=\textwidth]{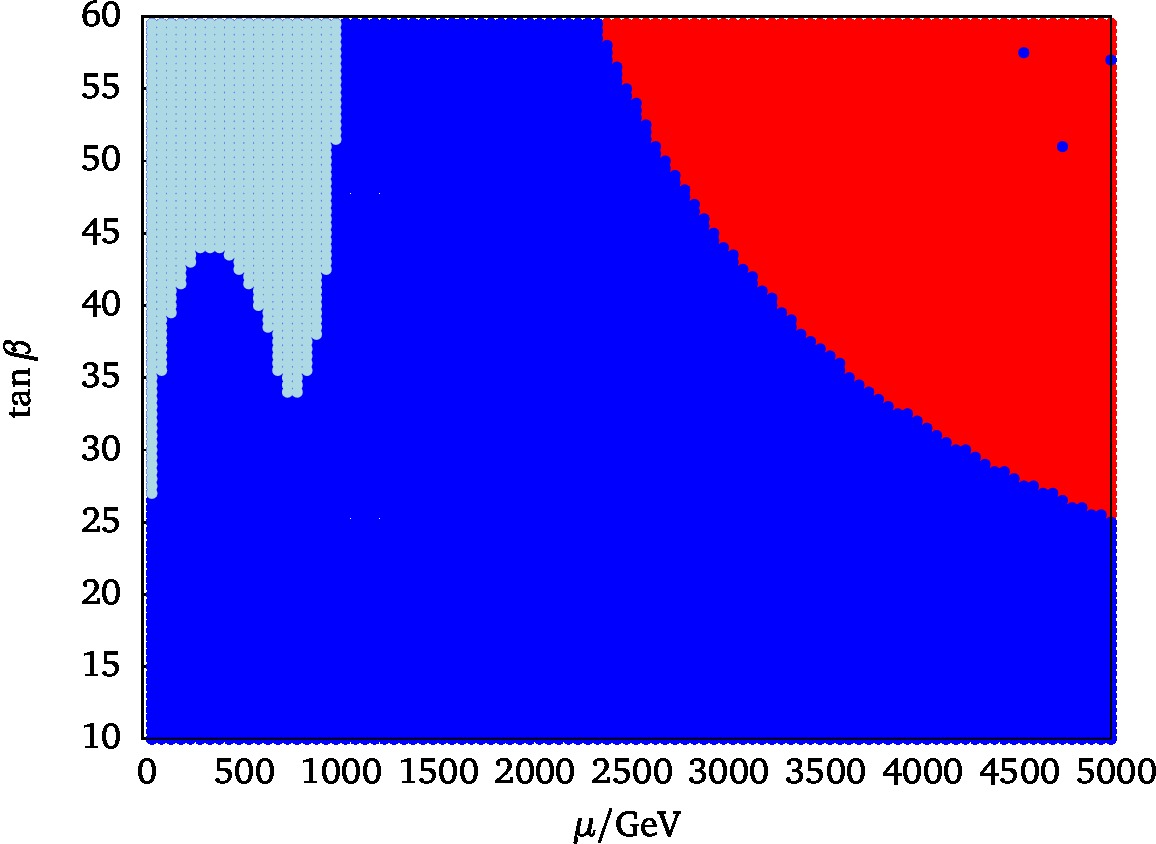}
\end{minipage}%
\begin{minipage}{0.48\textwidth}
\includegraphics[width=\textwidth]{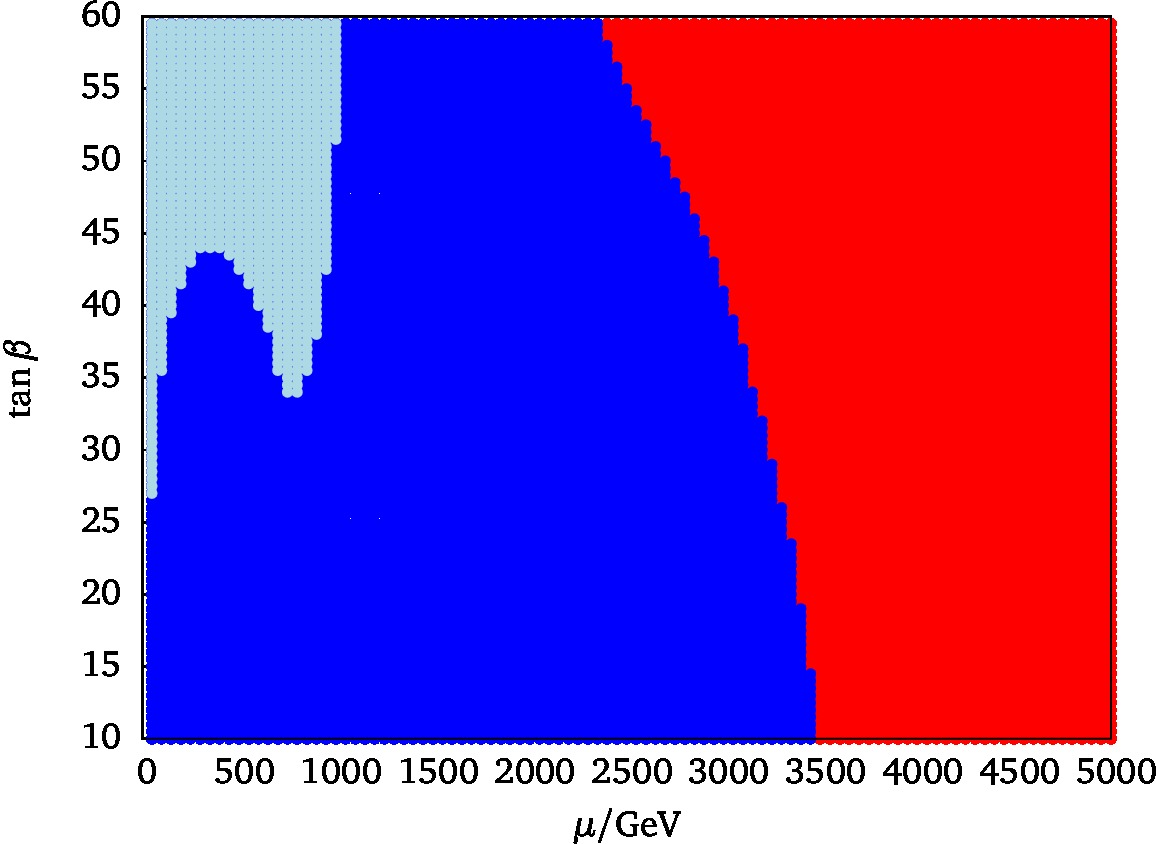}
\end{minipage}\\
\begin{minipage}{0.48\textwidth}
\includegraphics[width=\textwidth]{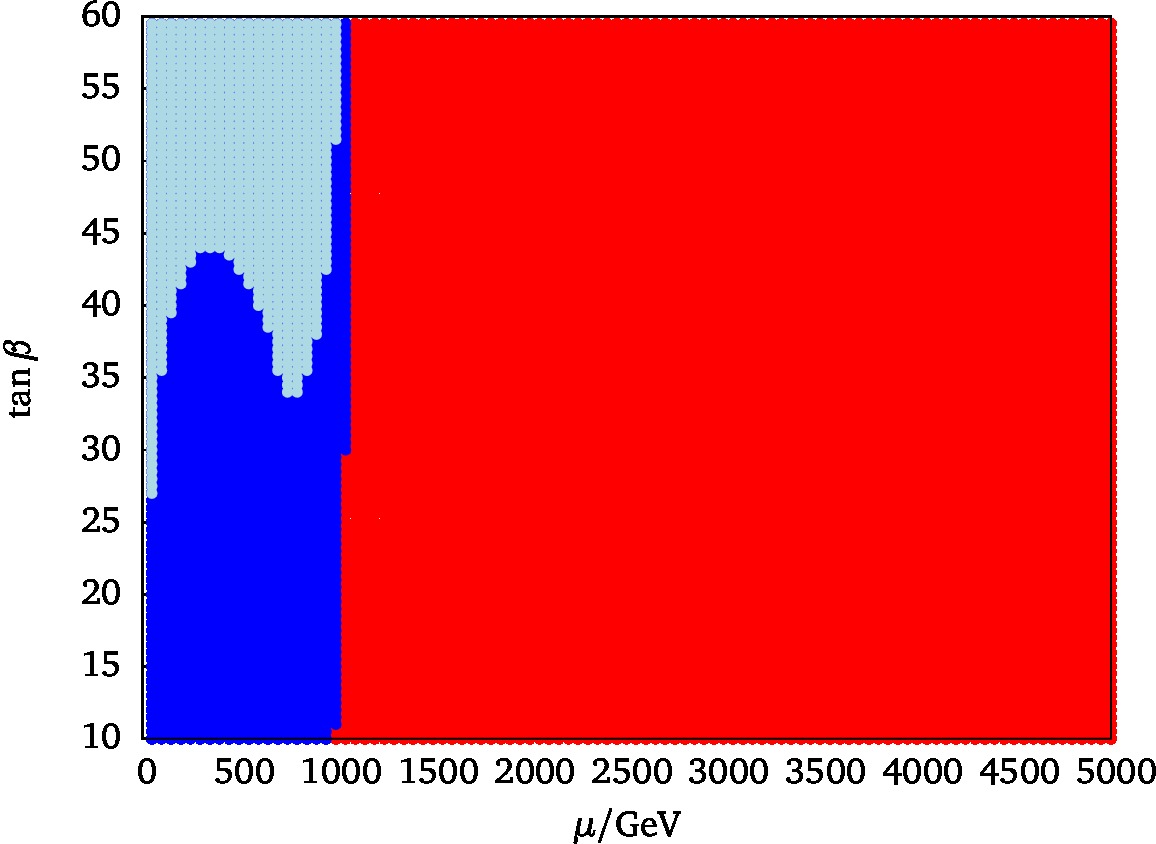}
\end{minipage}%
\begin{minipage}{0.48\textwidth}
\includegraphics[width=\textwidth]{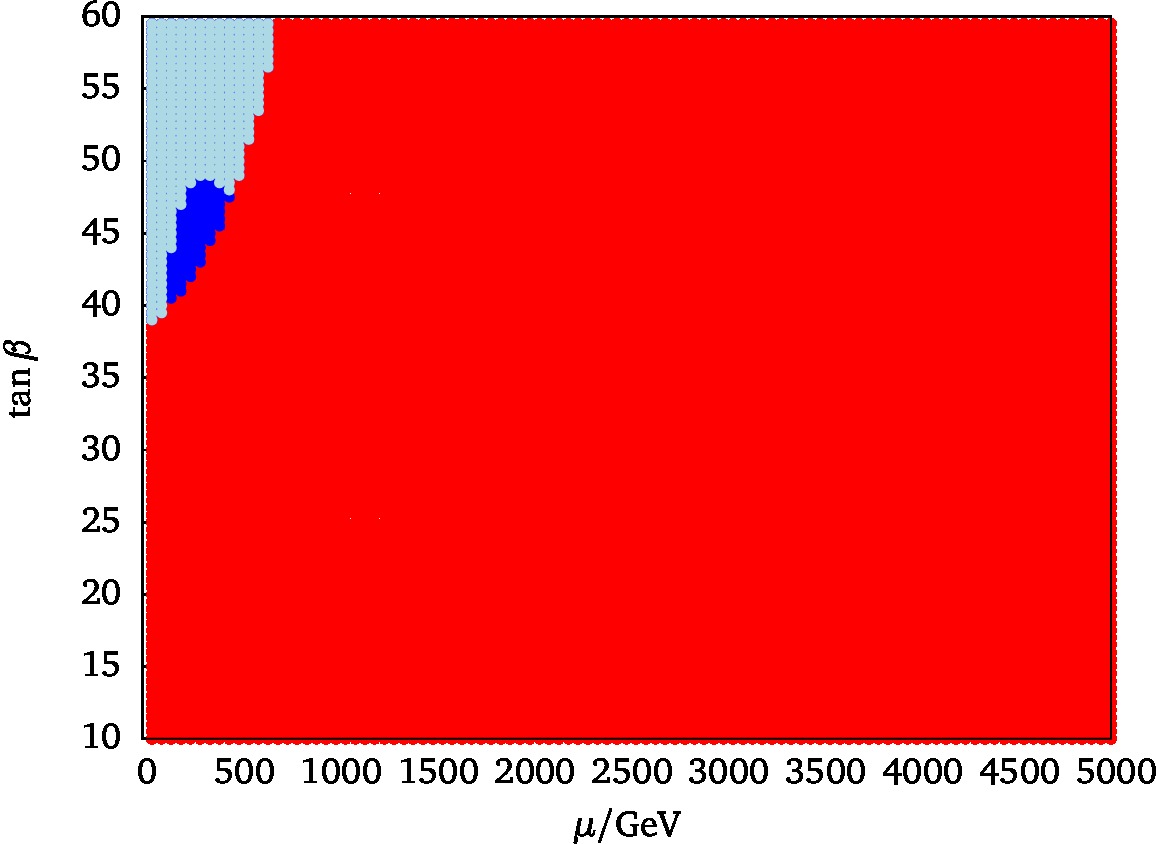}
\end{minipage}
\caption{Growing exclusion limits if more \vev{}s are allowed; red
  points are excluded by vacuum stability, blue points are still
  allowed. The light blue points indicate a light Higgs mass within
  \(125\pm 1\,\GeV\) (\textsc{FeynHiggs} with a gluino mass at \(3
  M_\text{SUSY}\)). All plots have \(A_\tq = -1500\,\GeV\) and
  \(M_\text{SUSY} = 1\,\TeV\) for comparison with previous works. Top
  left: similar configuration as in~\cite{Bobrowski:2014dla,
    Hollik:2015pra} with no stop and \(h_\dq\) \vev{}s; top right:
  including also \(h_\dq \neq 0\), where the exclusion is now
  interpolating between the two scenarios of~\cite{Hollik:2015pra} and
  is a bit stronger (as not necessarily \(|h_\dq|^2 = |h_\uq|^2 +
  |\tilde b|^2\) is fixed). Down left: including all field directions
  discussed in this paper, as in the upper row we kept \(A_\bq = 0\);
  down right: now switching on \(A_\bq = A_\tq\), nearly the complete
  area seems to be excluded. To compare with the usual (p)MSSM
  literature, we have to rescale the \(A\)-terms \(A_\tq \to A_\tq /
  y_\tq\) and \(A_\bq \to A_\bq / y_\bq\). In this area, \(y_\bq\)
  ranges from \(\sim 0.12\) to \(\sim 0.8\) and gets large in the upper
  left corner of the \(\mu\)-\(\tan\beta\) plane including the
  \(\Delta_b\) resummation but less large than \(m_\bq / v_\dq\) (which
  seems to rescue this corner once \(A_\bq\) is switched on).}
\label{fig:exclusions_mu-tb}
\hrule
\end{figure}

In comparison with earlier work on the vacuum stability issue in the
MSSM, we are now able to exclude a wider region of parameter space which
gets accessed when \emph{both} stop and sbottom \vev{}s and non-standard
values for the two Higgs doublets are considered. Non-standard in this
respect means \(\langle h_\uq \rangle \neq v_\uq = v \sin\beta\) and
\(\langle h_\dq \rangle \neq v_\dq = v \cos\beta\). Excluded regions in
the \(\mu\)-\(\tan\beta\) plane have been derived from an analysis of
the one-loop Higgs potential in the MSSM, where loop effects of third
generation sfermions have been included~\cite{Bobrowski:2014dla}. An
additional minimum seems to appear at a larger field value \(h_\uq\)
which is driven by the \(\mu\)-term and therefore the requirement is
that this non-standard (apparently charge and color conserving!) \vev{}
does not lead to a minimal value of the potential that is lower than at
the electroweak \vev{}. Actually, this behavior is an artifact of
neglecting colored directions in the potential already at the tree-level
leading to an imaginary part in the example of
Ref.~\cite{Bobrowski:2014dla} that was not understood (and therefore
just ignored). As this imaginary part is related to a tachyonic sbottom
mass at the new \vev{}, this indicates a CCB global minimum, where the
``one-loop global minimum'' is rather a saddle point of the potential in
the Higgs--sbottom field configuration. For the same configuration
(basically \(h_\dq = 0\) and \(\tilde b = h_\uq\)) the shape of the
exclusion is very much the same but a bit tighter and shown in the upper
left plot of Fig.~\ref{fig:exclusions_mu-tb}. The choice of \(h_\dq =
0\) basically follows from the consideration that if \(\langle h_\dq
\rangle = v_\dq\) kept fixed, this value can be neglected for large
\(\tan\beta\) with respect to the much larger \(\langle h_\uq \rangle >
v_\uq\). However, this choice (as well as \(\tilde b \sim h_\uq\)) does
not resemble the true behavior of the potential as can be seen, when
both \(h_\dq\) and \(\tilde b\) are treated as independent dynamical
variables as described above. If one commits to the genuine \(D\)-flat
direction only (say \(|h_\dq|^2 = |h_\uq|^2 + |\tilde b|^2\)), similarly
wrong exclusions (conclusions?) can be drawn. The comparison of these
two choices has been elaborated in~\cite{Hollik:2015pra} together with
the corresponding analytic bound Uneqs.~\eqref{eq:ccbbot}
and~\eqref{eq:ccbbot2}. The combined exclusion limit interpolates
between the two and is shown in the upper right plot of
Fig.~\ref{fig:exclusions_mu-tb}. Again, the artificial constraint
\(\tilde t = 0\) leads to weaker exclusions than under a non-vanishing
stop \vev{}. This behavior finally is shown in the lower left plot of
Fig.~\ref{fig:exclusions_mu-tb} and excludes large parts of the
\(\mu\)-\(\tan\beta\) plane. So far, we also have kept the trilinear
soft SUSY breaking sbottom parameter \(A_\bq\) to zero and employed a
large but moderate \(A_\tq = -1500\,\GeV\) (the negative sign was chosen
to enhance the sbottom-\vev{} bound related to \(y_\bq\)). As a
non-vanishing \(A_\bq\) drives the formation of vacua with \(\langle
\tilde b\rangle \neq 0\) and similarly \(\langle h_\dq \rangle \gg
v_\dq\), we close the remaining allowed parameter space to values of
\(\tan\beta \gtrsim 40\) and \(\mu \lesssim 700\,\GeV\) (in a world with
\({\tilde m}_Q = {\tilde m}_t = {\tilde m}_b = M_\text{SUSY} = 1\,\TeV\)
and \(m_A = 800\,\GeV\) as the relevant further input parameters).

%%%%%%%%%% FIGURE %%%%%%%%%%
%%%%%%%%%% % mu-tanb exclusion %%%
%%%%%%%%%% FIGURE %%%%%%%%%%

\begin{figure}
\begin{minipage}{0.48\textwidth}
\includegraphics[width=\textwidth]{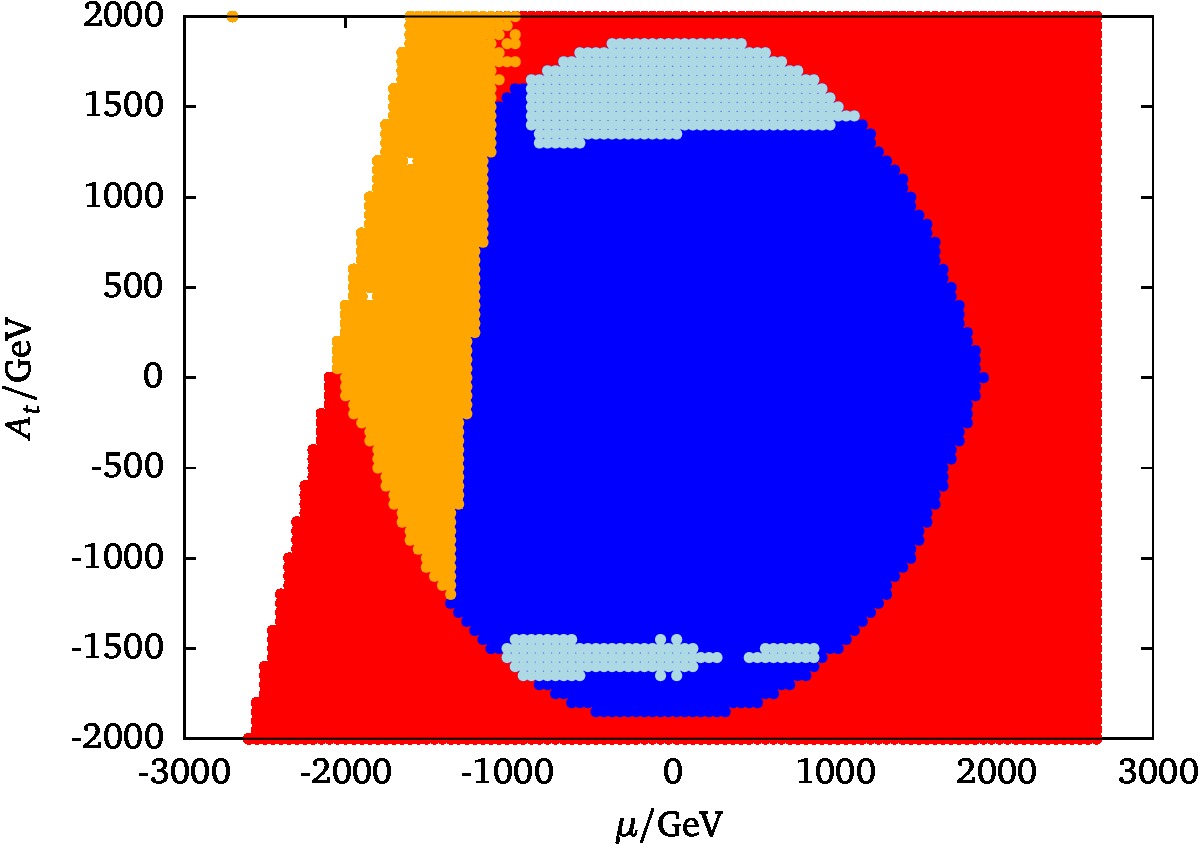}
\end{minipage}%
\begin{minipage}{0.48\textwidth}
\includegraphics[width=\textwidth]{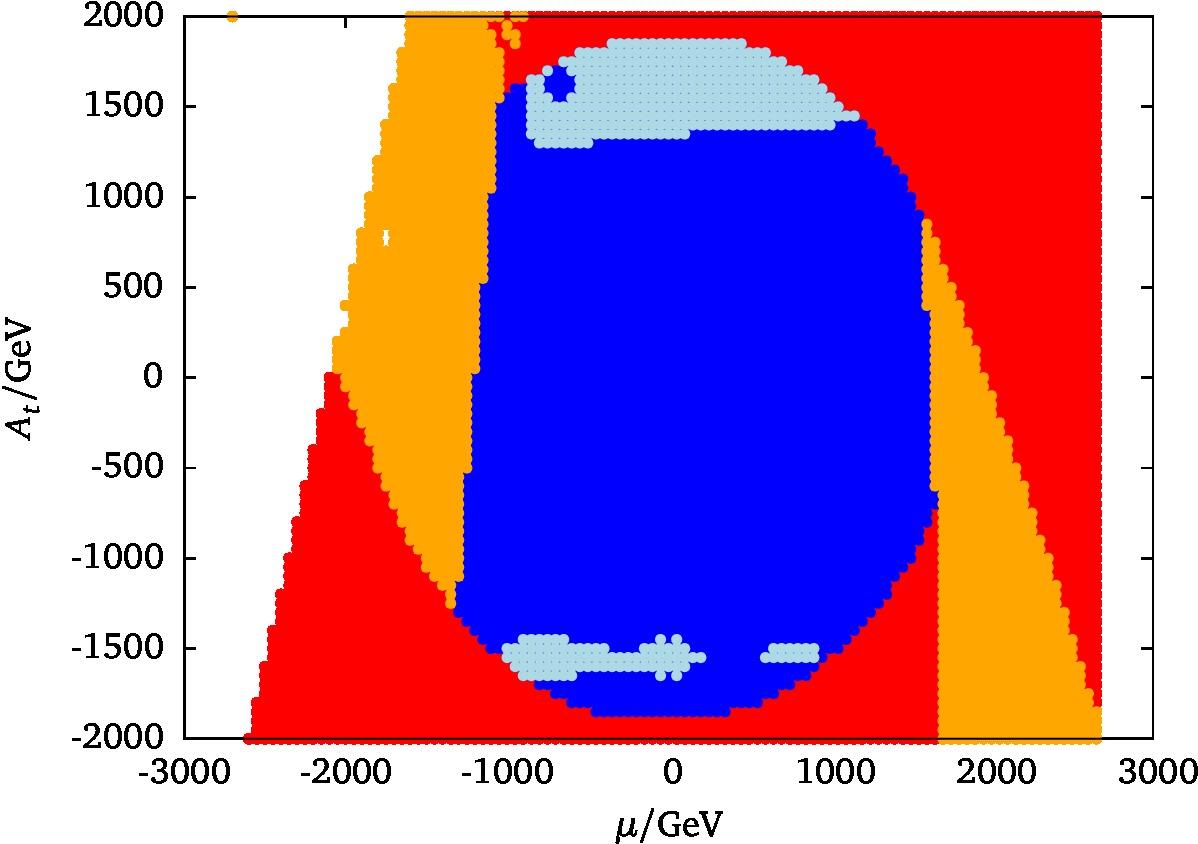}
\end{minipage}\\
\begin{minipage}{0.48\textwidth}
\includegraphics[width=\textwidth]{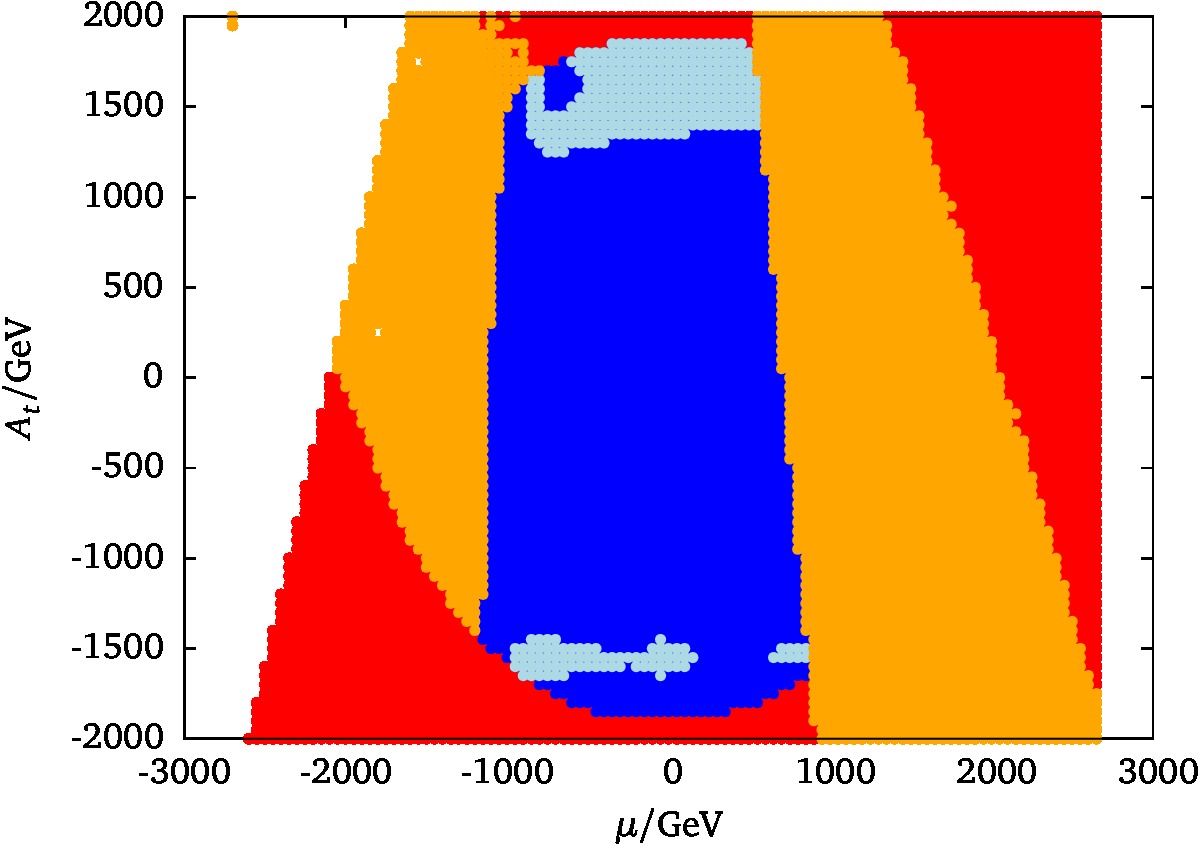}
\end{minipage}%
\begin{minipage}{0.48\textwidth}
\includegraphics[width=\textwidth]{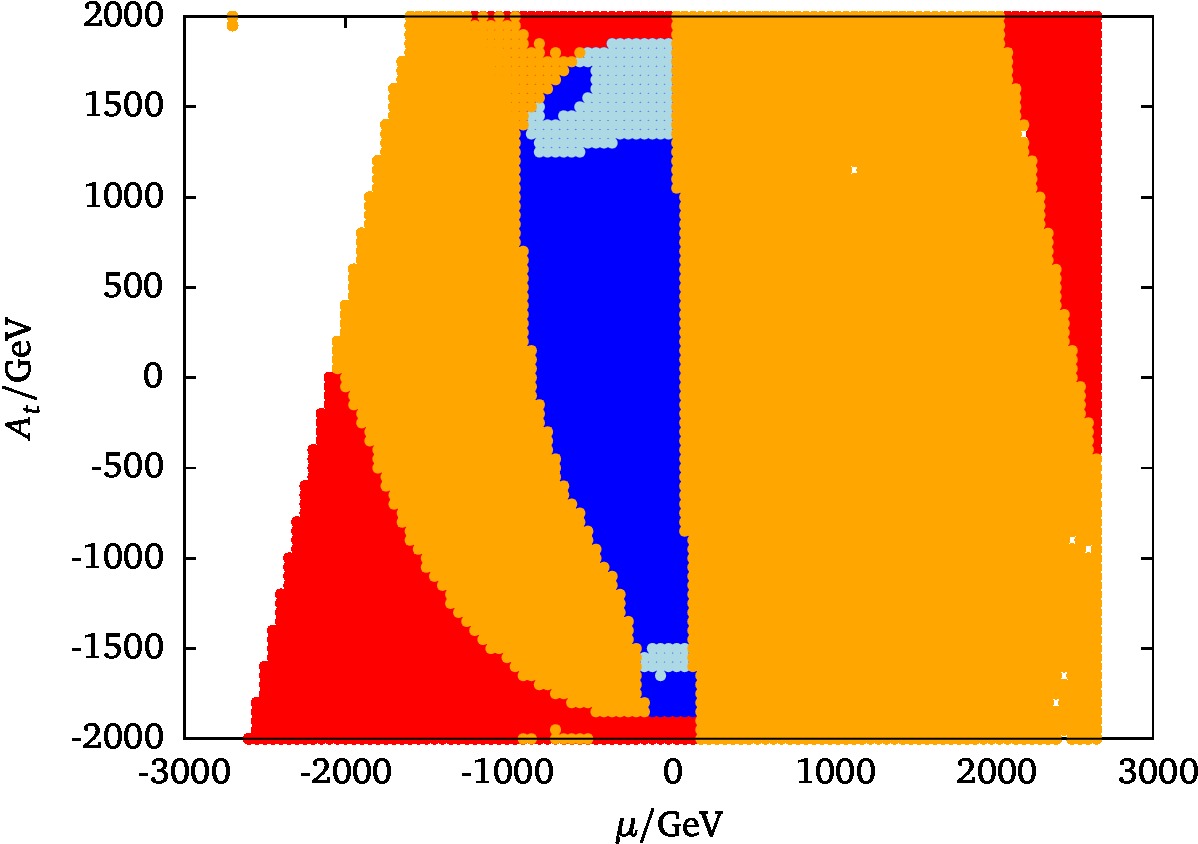}
\end{minipage}
\caption{We show exclusion limits from the formation of non-standard
  vacua in the \(A_\tq\)-\(\mu\) plane. The central blue area appears to
  be allowed, where the red and orange regions are excluded by the
  existence of stop and sbottom \vev{}s, respectively. Light blue points
  indicate the region of the correct light Higgs mass as in
  Fig.~\ref{fig:exclusions_mu-tb}. In all plots, we assigned
  \(M_\text{SUSY} = 1\,\TeV\). The soft SUSY breaking trilinear coupling
  is set to \(A_\bq = 0\,\GeV\) (upper left), \(A_\bq = 500\,\GeV\)
  (upper right), \(A_\bq = 1000\,\GeV\) (lower left) and \(A_\bq =
  1500\,\GeV\) (lower right).}
\label{fig:exclusions_At-mu}
\hrule
\end{figure}

How much is the interplay of stop and sbottom \vev{}s? The exclusion
from stop \vev{}s only has a rather circular shape in the
\(\mu\)-\(A_\tq\) plane, illustrated in
Fig.\ref{fig:exclusions_At-mu}. This shape does not change much with
\(\tan\beta\) as long as \(A_\bq\) is switched off. In the upper left
corner, we show exactly this for \(\tan\beta = 40\) and \(A_\bq =
0\). For the purposes of Figs.~\ref{fig:exclusions_mu-tb}
and~\ref{fig:exclusions_At-mu}, we relied on our own determination of
the Higgs potential parameters (basically \(m_{H_\uq}^2\) and
\(m_{H_\dq}^2\) to have the correct \(v_{\dq, \uq}\) in presence of the
one-loop corrected potential). Comparison of public codes doing the same
(\textsc{SPheno}~\cite{Porod:2003um, Porod:2011nf},
\textsc{softsusy}~\cite{Allanach:2001kg} and
\textsc{SuSpect}~\cite{Djouadi:2002ze} with the convenient
\textsc{Mathematica} interface SLAM~\cite{Marquard:2013ita}) shows very
similar shapes, where the border lines get less sharp due to several
effects we do not have under control. For aesthetic reasons, we show the
(slightly wrong but nicer) plots determined with our own algorithm. The
color coding in Fig.~\ref{fig:exclusions_At-mu} shows allowed regions in
blue, excluded by stop \vev{}s in red and sbottom \vev{} appearing in
orange. As can be seen by turning on \(A_\bq\), a larger portion of the
previously allowed parameter space is excluded. The allowed parameter
space for \(m_{h^0}\) within a \(1\,\GeV\) interval around \(125\,\GeV\)
is shown in light blue.

%%%%%%%%%% FIGURE %%%%%%%%%%
%%%%%%%%%% % mu-At exclusions %%%
%%%%%%%%%% FIGURE %%%%%%%%%%

So far, we only analyzed the very generic potential of
Eq.~\eqref{eq:hisqpot}, rewritten as single-field
potential~\eqref{eq:onefieldpot}, without any reference to current
phenomenology of the MSSM. It appears that the pure theoretical
consideration to have a self-consistent theory (especially having no
deeper minimum than the electroweak ground state) already excludes wide
regions of the available parameter space. The constraints are even
stronger than the well-known strong constraints
of~\cite{Casas:1995pd}. Reasons are that we do not insist on \(\tan\beta
> 1\) for the new vacuum and include simultaneously stop and sbottom
\vev{}s. Unfortunately, for direct comparison with the
\textsc{Vevacious} collection, the corresponding model file treating
non-vanishing stop and sbottom \vev{}s at the same time without the need
of the \emph{full} squark potential including the first two generations
is missing (although there \(|h_\dq| > |h_\uq|\) is allowed and field
values can also acquire negative values with respect to \(h_\uq\) as we
do; the usage of the full squark potential appears to be less stable and
requires very long running times for each data point). In its full
generality, however, \textsc{Vevacious} is not constrained to the MSSM
and can be used to check the stability of any desired beyond the SM
physics---if the user is willing to produce the necessary model file.
Even more constrained gets the leftover parameter space when we in
addition impose the light Higgs mass \(m_{h^0} = 125\,\GeV\) which was
not available 20 years ago and on its own narrows down the allowed
region. Of course, a detailed analysis of the MSSM parameter space can
only be done in terms of a global fit including various other
constraints (collider data, Higgs properties, dark matter constraints)
and goes much beyond the scope of this work. The analysis of the CCB
anatomy, however, is interesting by itself and even more in connection
to the determination of the Higgs mass.

\begin{figure}
\begin{minipage}{0.48\textwidth}
\includegraphics[width=\textwidth]{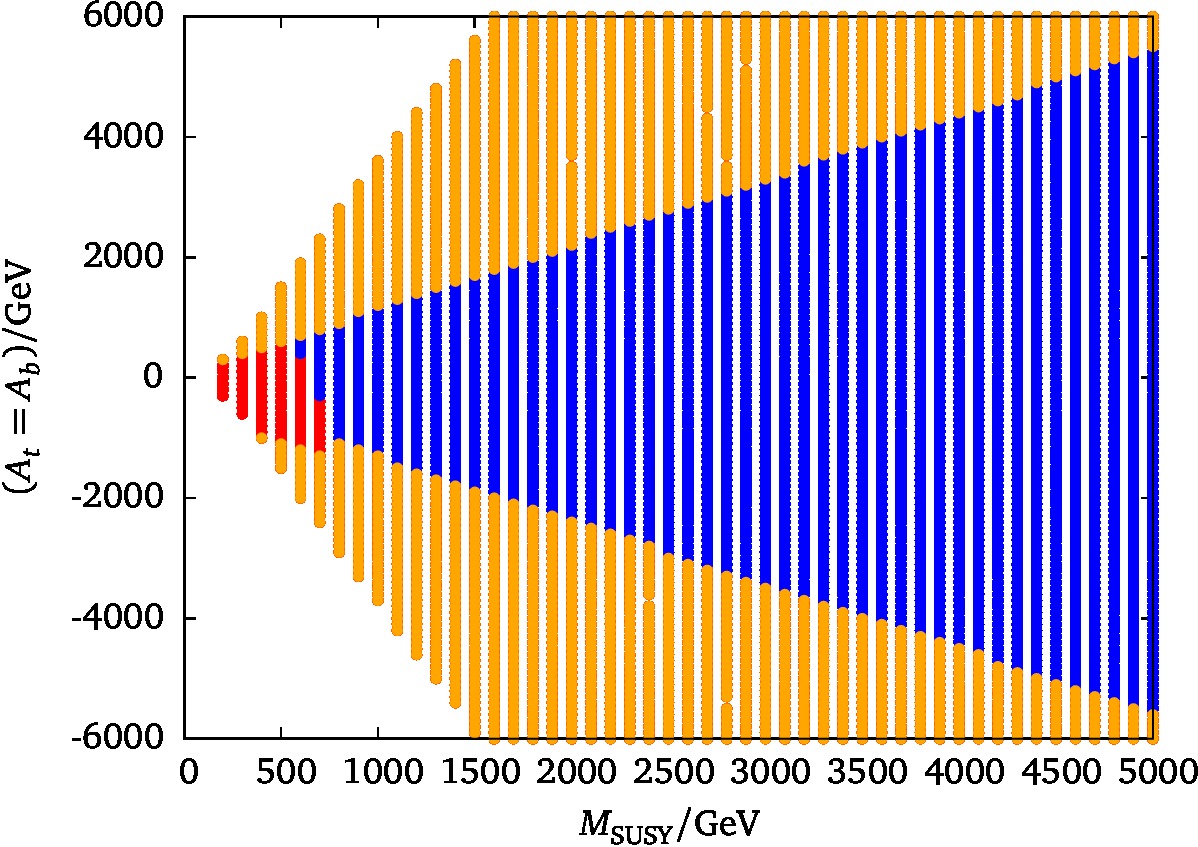}
\end{minipage}
\begin{minipage}{0.48\textwidth}
\includegraphics[width=\textwidth]{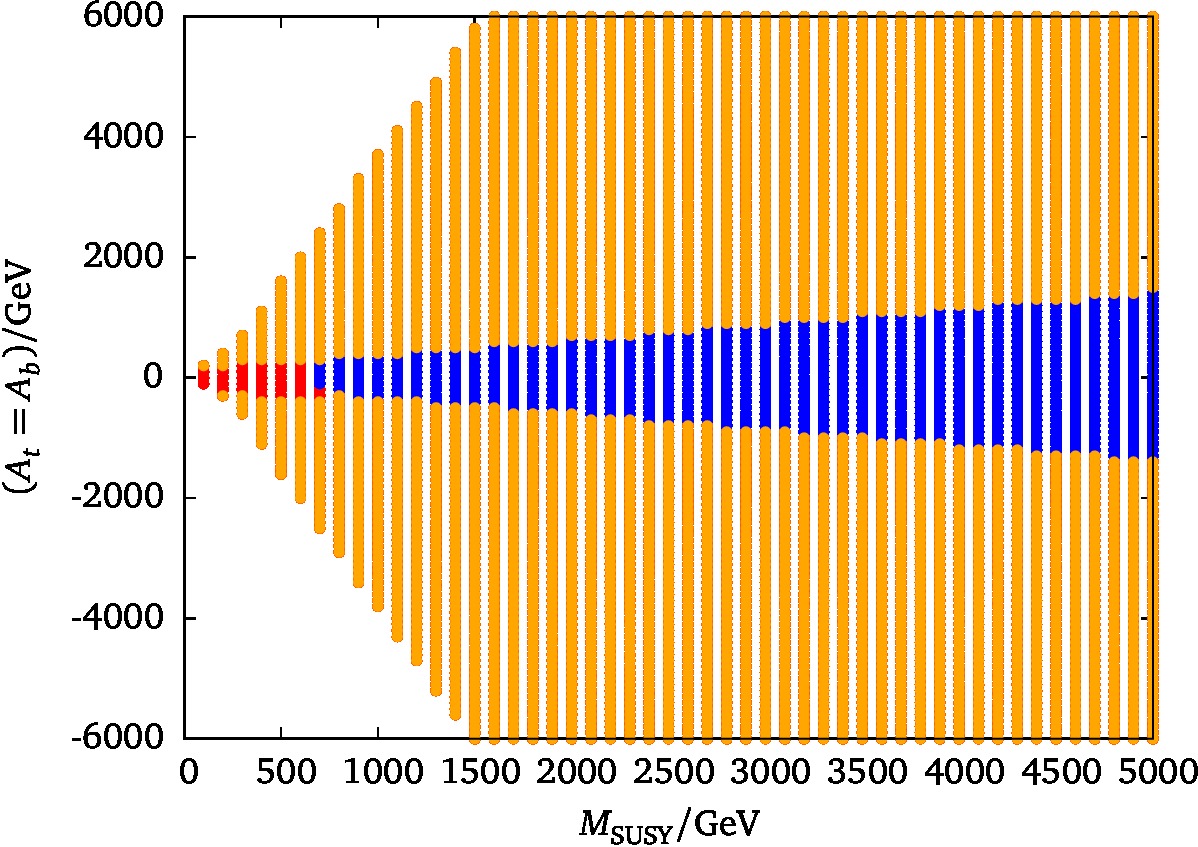}
\end{minipage}
\caption{The allowed and excluded regions for varying SUSY masses
  \({\tilde m}_Q = {\tilde m}_t = M_\text{SUSY}\) and \(\tan\beta = 40\)
  (left) as well as \(\tan\beta = 10\) (right) and \(\mu = 350\,\GeV\)
  (both). Blue points do not show any deeper non-standard vacua whereas
  red and orange do. Orange points have explicitly non-vanishing sbottom
  \vev{}s. For the purpose of these plots, we have employed
  \textsc{SPheno} to calculate the spectra.}\label{fig:At-MSUSY}
\hrule
\end{figure}

\begin{figure}
\begin{minipage}{0.48\textwidth}
\includegraphics[width=\textwidth]{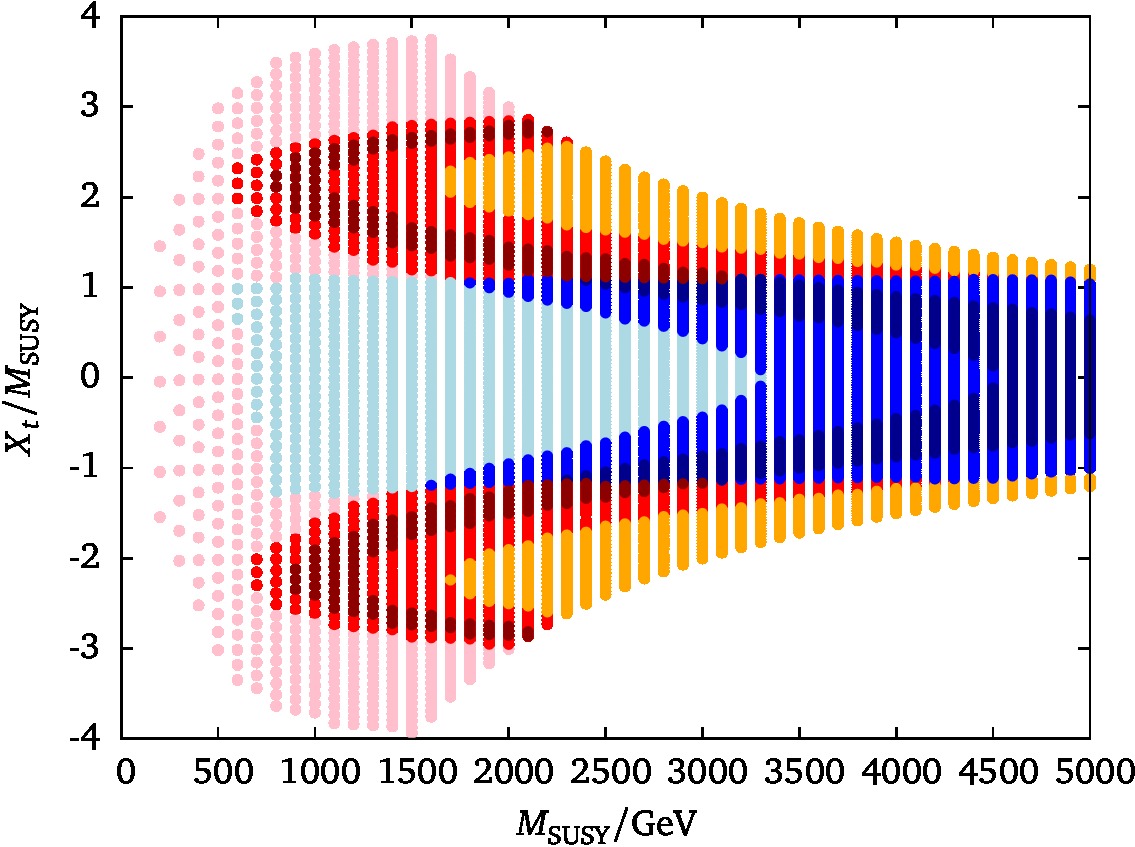}
\end{minipage}
\begin{minipage}{0.48\textwidth}
\includegraphics[width=\textwidth]{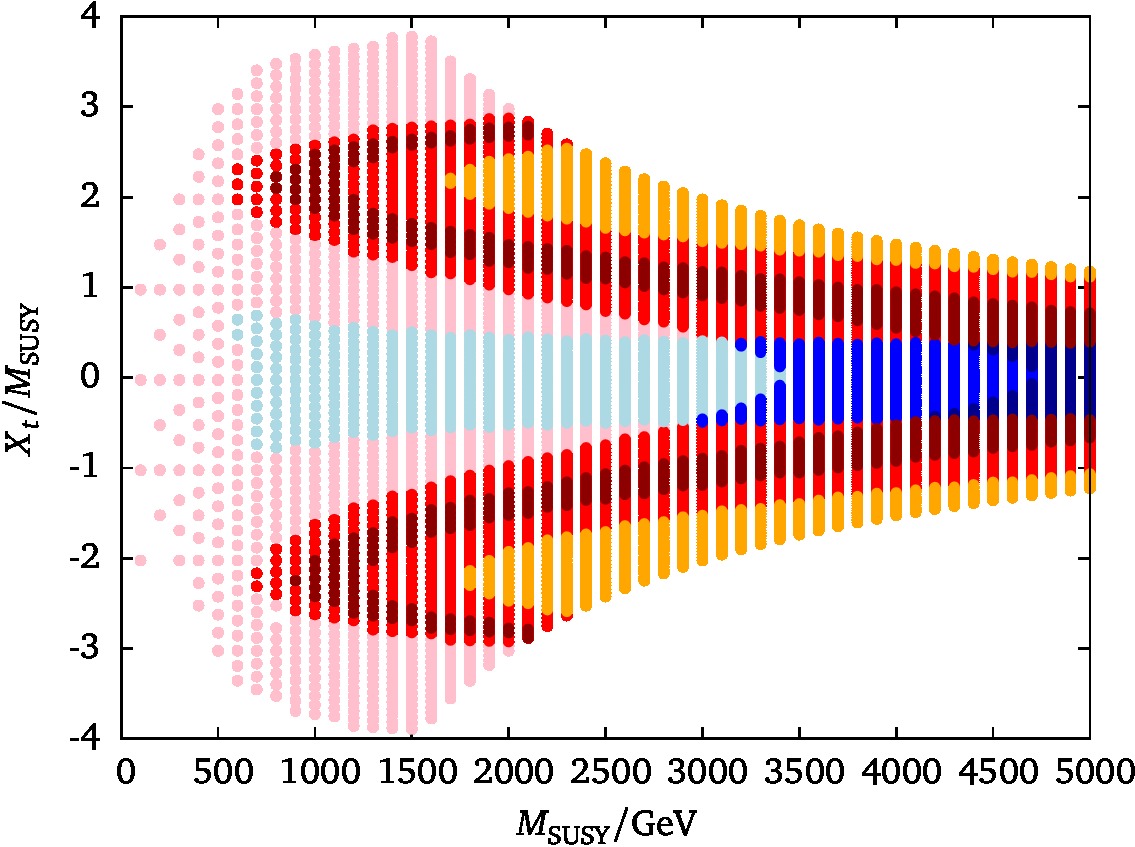}
\end{minipage}
\caption{For the inclusion of the Higgs mass prediction in the MSSM, we
  show the value of \(X_t / M_\text{SUSY}\) varying with the SUSY scale
  (at \(A_\bq = A_\tq\) as usually done). On the left side, we kept a
  low \(\mu\)-value, \(\mu = 350\,\GeV\), fixed; whereas on the right
  side, \(\mu\) scales with the SUSY mass, \(\mu = M_\text{SUSY}\). In
  both cases we employed \(\tan\beta = 40\). The color coding shows the
  compatibility with the Higgs mass measurements: the darker the color
  the more compatible (the dark blue and red regions indicate \(m_{h^0}
  \in [124, 126]\,\GeV\) as produced by \textsc{SPheno} corresponding to
  a \(1\,\GeV\) uncertainty; the neighboring region has the \(3\,\GeV\)
  uncertainty). The blueish region is allowed by vacuum stability
  considerations whereas the reddish region is excluded. For fixed \(\mu
  = 350\,\GeV\) that means roughly \(|X_t| \leq 1.1 M_\text{SUSY}\); in
  the case with \(\mu = M_\text{SUSY}\) the allowed area shrinks with
  increasing SUSY scale.}\label{fig:Xt-MSUSY}
\hrule
\end{figure}

\paragraph{SUSY hides behind the corner}
The question that remains is how much do these bounds depend on the SUSY
scale. So far, we have employed \(M_\text{SUSY} = 1\,\TeV\) which needs
large \(A_\tq\) close to the border line in order to get \(m_{h^0}\)
right and what points towards near-criticality also in the MSSM. There
is of course one way out to still consider the MSSM (with the
assumptions applied in this work) as valid and alive. We find that the
constraints get weaker with increasing \(M_\text{SUSY}\), especially the
value of the ratio \(X_t / M_\text{SUSY}\) for which a model point would
be excluded stays rather constant or even grows for fixed \(\mu\)
whereas it shrinks with larger \(\mu\). This shrinking is not
surprising, as \(\mu\) enters \(X_t\) and can enhance its value for
various configurations. Similarly, the value of the tree-level
\(\tan\beta\) enters severely as can be seen from
Fig.~\ref{fig:At-MSUSY}, where we compare the allowed and excluded
regions of \(A_\tq = A_\bq\) with respect to \(M_\text{SUSY}\) for
\(\tan\beta = 40\) and \(10\).. In addition, the proper value for the
lightest Higgs mass, \(m_{h^0} \approx 125\,\GeV\) selects a small band
in \(X_t\)-\(M_\text{SUSY}\). We clearly see from
Fig.~\ref{fig:Xt-MSUSY}, where the spectrum has been determined with the
help of \textsc{SPheno}, that only for SUSY masses that are anyway not
yet excluded by experiment in the simplified analyses \(M_\text{SUSY}
\geq 1500\,\GeV\) (for a small \(\mu\)-term and rather large
\(\tan\beta\)), we can enter the correct range. Increasing \(\mu\)
shifts the allowed regime to even larger \(M_\text{SUSY}\). It is
therefore with hindsight not surprising at all, that there have been no
signals of SUSY found so far in combination to the measured light Higgs
mass. Without this additional crucial ingredient one might get depressed
seeing the parameter space being closed, especially when the trilinear
soft SUSY breaking couplings are taken equally large, \(A_\bq = A_\tq\),
as usually done. One the other hand, this is exactly what is observed by
the non-observation of light stops so far. Very light squarks (below say
\(1\,\TeV\)) in connection with large squark mixing are inconsistent
with a stable electroweak vacuum. In that sense, SUSY awaits her
discovery in the very near future.

\section{Conclusions}
We have reported on a new view of charge and color breaking minima in
the Minimal Supersymmetric Standard Model and consequently derived novel
bounds on the parameter space from the self-consistency of the
theory. In order to avoid any configurations that lead to an unstable
electroweak ground state, large portions of the available parameter
space are excluded. We have argued that the exclusions cannot be treated
as metastability bounds requiring only a life-time of the false vacuum
of about the age of the universe and any CCB exclusion an MSSM
parameters is to be seen strict. We have extended the exhaustive work of
Ref.~\cite{Casas:1995pd} mostly by relaxing the constraint on \(h_\dq\)
to be strictly smaller than \(h_\uq\) but lacking simple analytic
expressions to cover the numerical exclusions. By analyzing both Higgs
and third generation squark directions simultaneously (four fields), we
cannot fix the signs of the trilinear terms to make them positive. This
in addition opens a new window to exclude larger parameter regions as
\(h_\dq = - h_\uq\) is allowed and particularly enhances the effect for
certain sign combinations. A generic analytic bound on the four-field
level is rather impossible; the remaining freedom, however, allows not
to be too restrictive and especially allow for short-lived vacua in
formerly metastable parameter regions.

\begin{figure}
\begin{minipage}{0.48\textwidth}
\includegraphics[width=\textwidth]{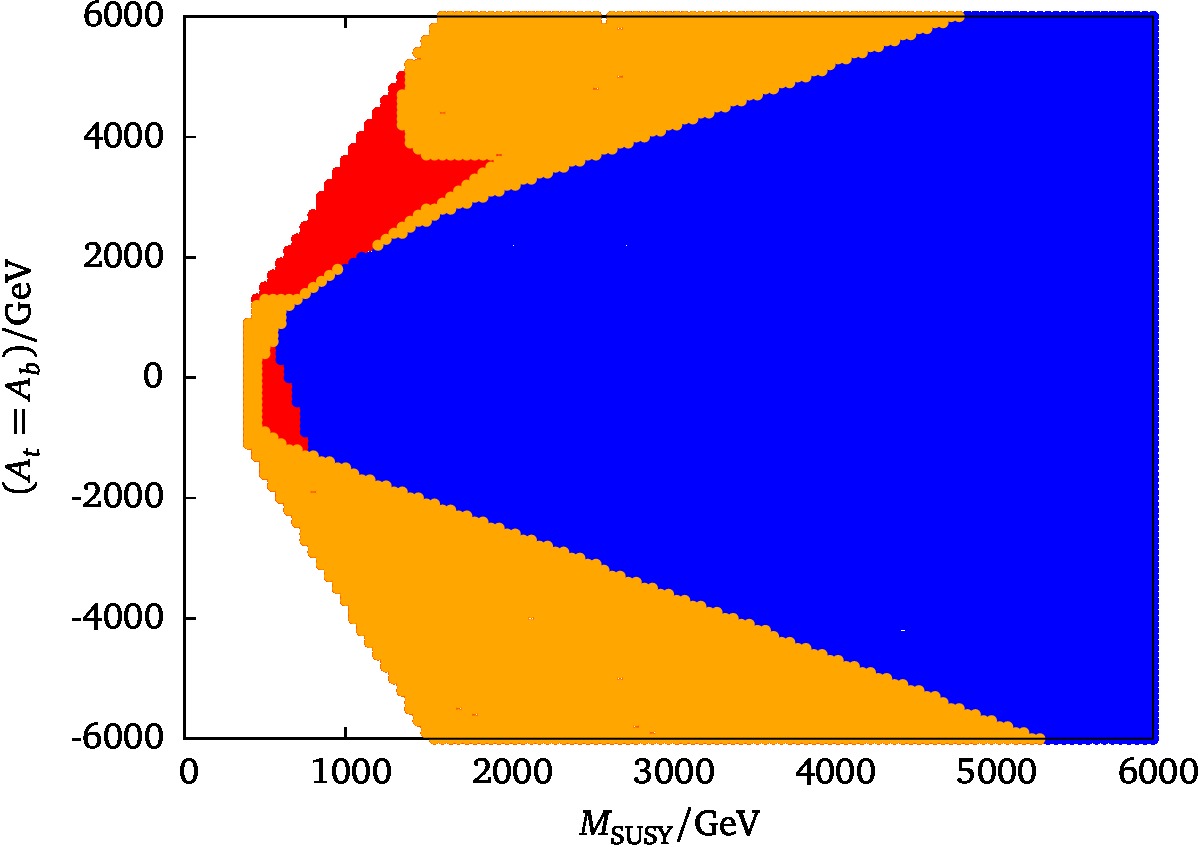}
\end{minipage}%
\begin{minipage}{0.48\textwidth}
\includegraphics[width=\textwidth]{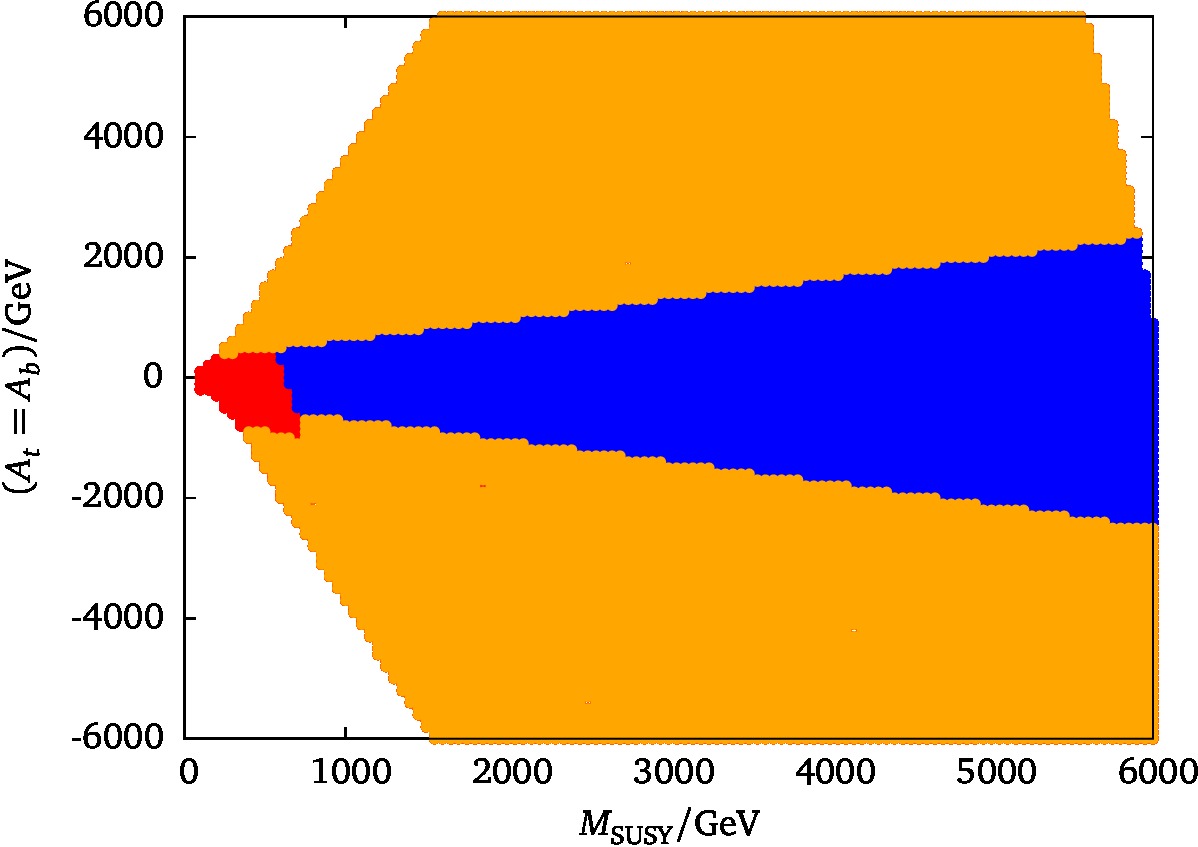}
\end{minipage}\\
\begin{minipage}{0.48\textwidth}
\includegraphics[width=\textwidth]{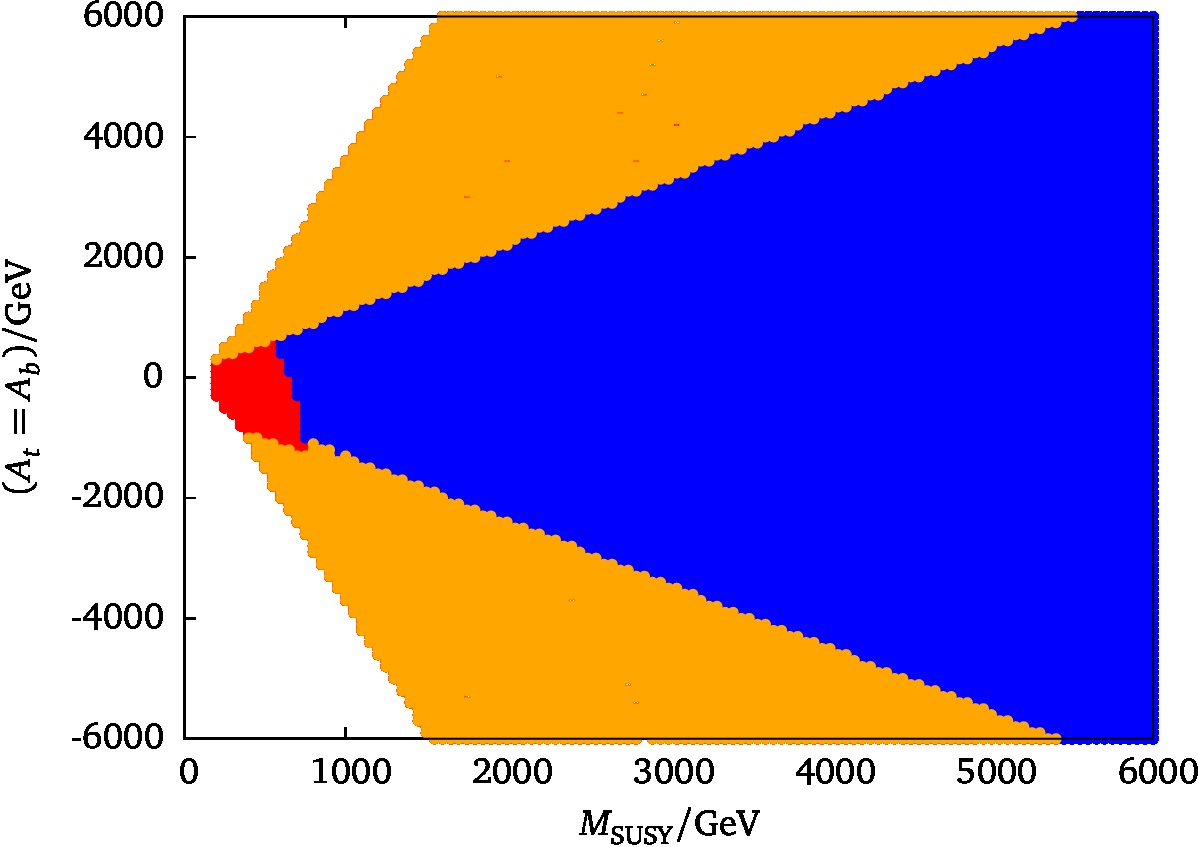}
\end{minipage}%
\begin{minipage}{0.48\textwidth}
\includegraphics[width=\textwidth]{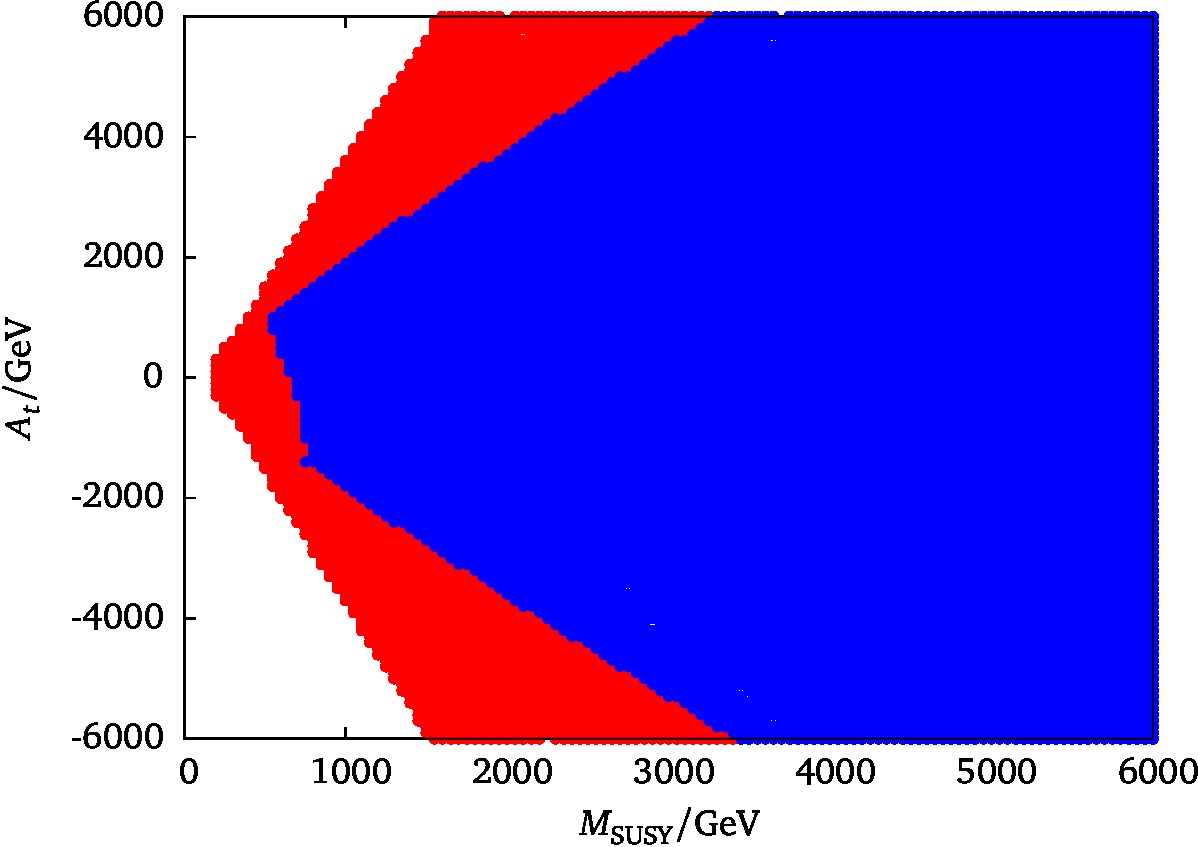}
\end{minipage}
\caption{Variations of Fig.~\ref{fig:At-MSUSY}, all with \(\tan\beta =
  40\) and \(m_A = 800\,\GeV\). All but the lower right have \(A_\bq =
  A_\tq\), in the lower right plot we have set \(A_\bq = 0\,\GeV\). For the
  upper left, we keep a negative \(\mu\)-value \(\mu = -500\,\GeV\)
  fixed, where the upper right has a varying \(\mu = M_\text{SUSY}\),
  where the lower left has \(\mu=350\,\GeV\) fixed. Color coding as in
  Fig.~\ref{fig:At-MSUSY}.}
\label{fig:At-MSUSY-var}
\hrule
\end{figure}

\begin{figure}
\begin{minipage}{0.48\textwidth}
\includegraphics[width=\textwidth]{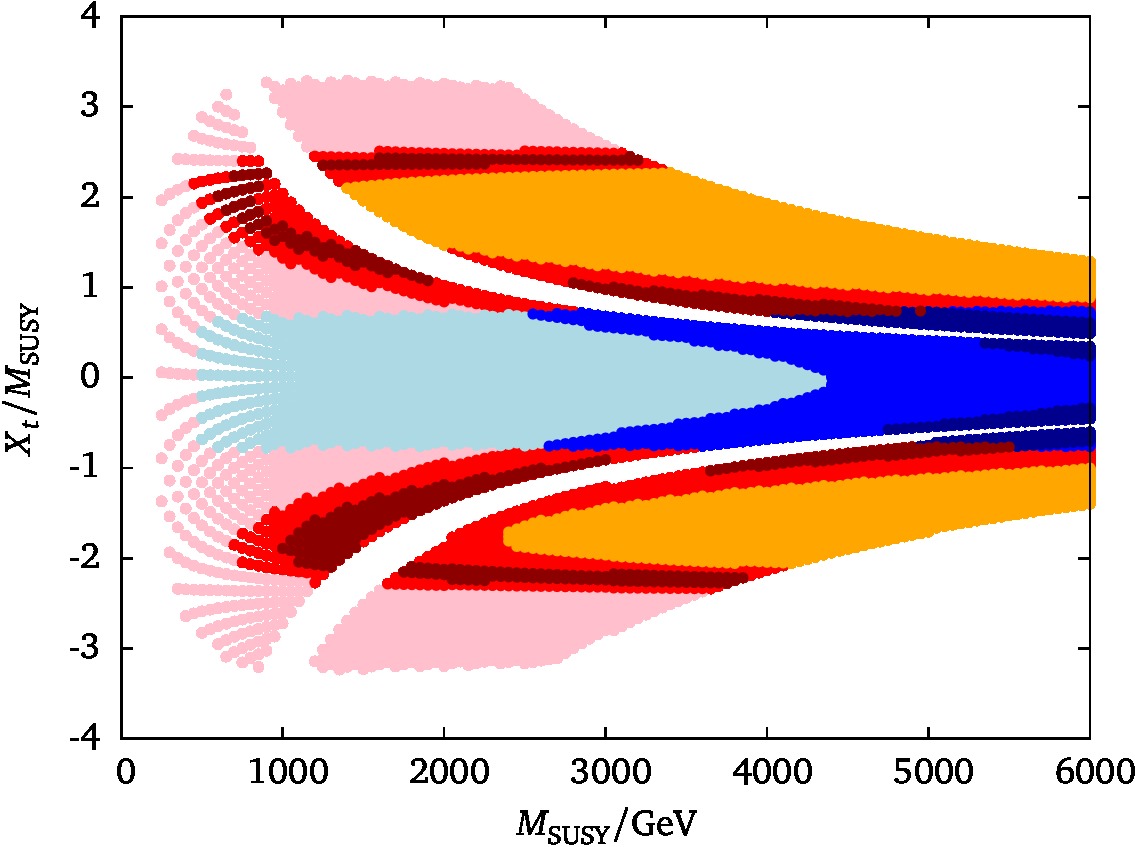}
\end{minipage}%
\begin{minipage}{0.48\textwidth}
\includegraphics[width=\textwidth]{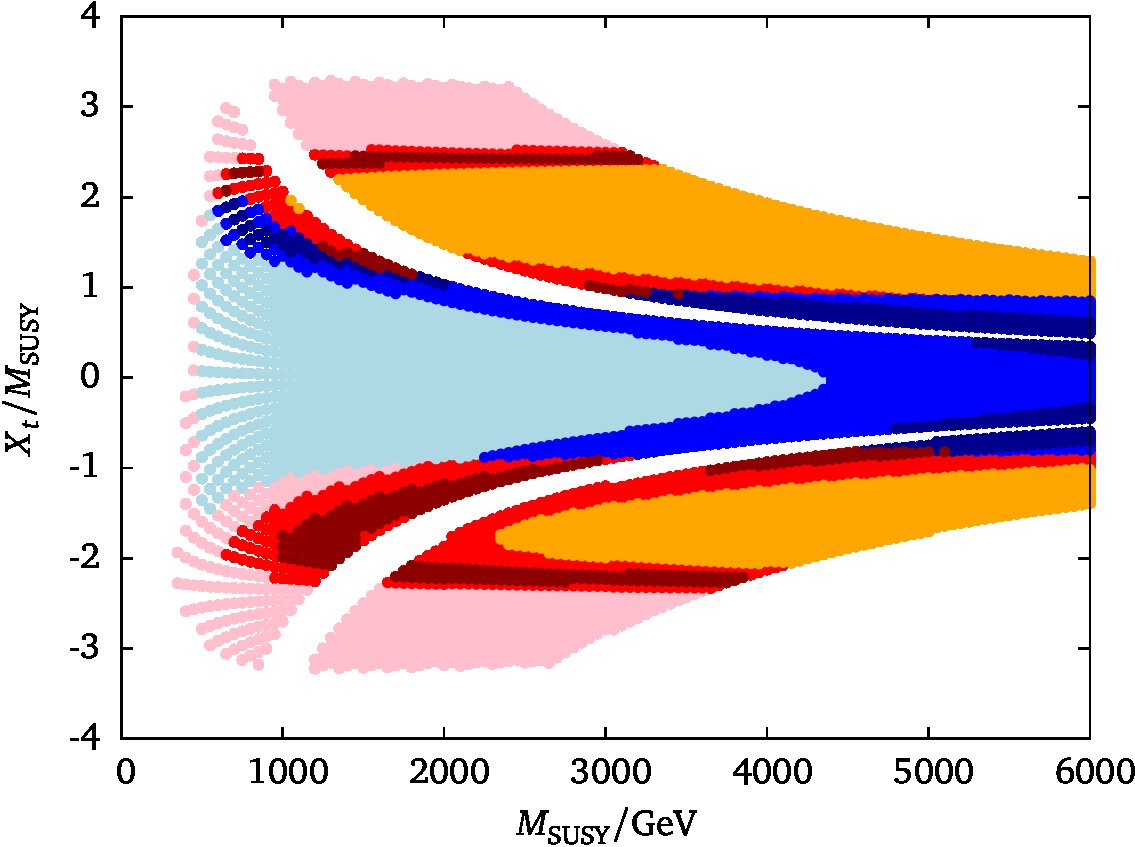}
\end{minipage}
\caption{Variations of Fig.~\ref{fig:Xt-MSUSY} with \(\tan\beta =
  40\) and \(\mu = 350\,\GeV\) (left) as well as \(\mu = -500\,\GeV\)
  (right). The light CP-even Higgs mass has been calculated with the
  help of \textsc{FeynHiggs} and \(m_{\tilde g} = 1.5 M_\text{SUSY}\),
  \(m_A = 800\,\GeV\). The white stripes are left blank because the
  one-loop effective potential of~\cite{Bobrowski:2014dla} develops an
  imaginary part already at the standard \vev{}s \emph{and} a tachyonic
  sbottom mass there.}
\label{fig:Xt-MSUSY-var}
\hrule
\end{figure}

Finally, we have included the determination of the light Higgs mass in
the MSSM and find that a low superpartner spectrum (especially light
stops) in combination with a \(125\,\GeV\) Higgs is excluded by the
formation of non-standard vacua around the SUSY scale. A stable
electroweak vacuum at the low scale requests (depending on the specific
scenario) SUSY masses to be in the multi-TeV regime, \(M_\text{SUSY}
\gtrsim 1.5,\dotsc,6\,\TeV\), for positive \(\mu\)-values and sizeable
\(A_\bq\). Exclusions get weakened for smaller or vanishing
\(A_\bq\). Variations on the bounds with rising \(M_\text{SUSY}\) are
given in Fig.~\ref{fig:At-MSUSY-var}. Further investigation is needed
in very special corners of the parameter space see
Fig.~\ref{fig:Xt-MSUSY-var}: \emph{negative} and small values of \(\mu\)
keep the window for a \(125\,\GeV\) Higgs and squark masses below
\(1\,\TeV\) open (say \(0 \geq \mu \geq -1000\,\GeV\), as already
indicated in Fig.~\ref{fig:exclusions_At-mu} for \(A_\tq \approx A_\bq
\approx \pm 1500\,\GeV\)).

\newpage
\section*{Acknowledgments}
The author acknowledges support by the DESY fellowship program and
discussion with E.~Bagnaschi, S.~Di Vita, S.~Passehr, and G.~Weiglein.
We thank F.~Staub for communications on \textsc{Vevacious}.

\singlespacing\small
\bibliography{Bibliography}
\end{document}